\documentclass[aps,prb,twocolumn,showpacs,superscriptaddress]{revtex4}
\usepackage{amsmath,amssymb,graphicx,bm}

\newcommand{\ii}{\mathrm{i}}

\newif\ifpdf
\ifx\pdfoutput\undefined
\pdffalse
\else
\pdfoutput=1
\pdftrue
\fi
\ifpdf
\usepackage{graphicx}
\usepackage{epstopdf}
\else
\usepackage{graphicx}
\fi

\def\const{\mathrm{const}\,}
\def\SZ{S_z^{\mathrm{tot}}}

\begin{document}

\title{Asymmetric spin-1/2 two-leg ladders}

\author{D. N. Aristov}
\affiliation{Petersburg Nuclear Physics
Institute, Gatchina  188300, Russia} 
\affiliation{ Institut f\"ur Nanotechnologie, Karlsruhe Institute of Technology,
 76021 Karlsruhe, Germany }
\author{C. Br\"unger }
\affiliation{Institut f\"ur Theoretische Physik, Universit\"at
W\"urzburg, D-97074 W\"urzburg, Germany}
\author{ F. F. Assaad}
\affiliation{Institut f\"ur Theoretische Physik, Universit\"at
W\"urzburg, D-97074 W\"urzburg, Germany}
\author{M. N. Kiselev}
\affiliation{The Abdus Salam International Centre for Theoretical
Physics, Strada Costiera 11, Trieste, Italy}
 \author{A. Weichselbaum}
\affiliation{Physics Department, Arnold Sommerfeld Center for Theoretical Physics and Center 
for NanoScience, Ludwig-Maximilians-Universit\"at, 80333 Munich, Germany }
\author{S. Capponi}
\affiliation{Laboratoire de Physique Th\'eorique, IRSAMC, Universit\'e
Paul Sabatier, CNRS, 31062 Toulouse, France}
\author{F. Alet}
\affiliation{Laboratoire de Physique Th\'eorique, IRSAMC, Universit\'e
Paul Sabatier, CNRS, 31062 Toulouse, France}
%

\begin{abstract}
We consider asymmetric spin-1/2 two-leg ladders with non-equal antiferromagnetic (AF)
couplings $J_{\|}$ and $\kappa J_{\|}$ along legs ($\kappa\le1$) 
and ferromagnetic rung coupling, $J_{\perp}$. This model 
is characterized by a gap $\Delta $ in the spectrum of spin excitations. 
We show that in the large $J_{\perp}$ limit this gap is equivalent to
the Haldane gap for the AF spin-1 chain, irrespective of the asymmetry of the ladder.   
The behavior of the gap at small rung coupling falls in two different universality classes. The first 
class, which is best understood from  the case of the conventional symmetric ladder at $\kappa=1$, admits a linear scaling for the spin gap 
$\Delta \sim J_{\perp}$. The second class appears for a strong asymmetry of the coupling along legs, 
$ \kappa J _{\|} \ll J_{\perp} \ll  J _{\|}$ and is characterized by two energy scales: 
the exponentially small spin gap  $\Delta \sim J_{\perp} \exp(-J _{\|}/J_{\perp})$, and 
the bandwidth of the low-lying excitations induced by a Suhl-Nakamura indirect exchange
$\sim J_{\perp}^{2} /J_{\|} $. We report numerical results obtained by exact diagonalization, 
density matrix renormalization group and quantum Monte Carlo simulations for the spin gap and various spin correlation functions. 
Our data indicate that the behavior of the string order parameter, characterizing the hidden AF order in Haldane phase, 
 is different in the limiting cases of weak and strong asymmetry. 
On the basis of the numerical data, we propose a low-energy theory of effective spin-1 variables, pertaining to 
large blocks on a decimated lattice. 
\end{abstract}
\maketitle
%


\section{Introduction}

Recent progress in nanotechnologies, molecular electronics 
and quantum computing reinvigorated the 
interest to low-dimensional systems. 
Special attention has focused during recent years on quantum dots, arrays  of coupled quantum dots, \cite{Lee2002}
quantum wires, spin chains or ladders.  
Another class of physical systems where low-dimensionality can be achieved is ultra-cold gases in optical lattices, 
\cite{Bloch2008,Giamarchi2010}
which form good prototype systems for investigation of many strongly-correlated effects, such as 
 metal-insulator transition, low-dimensional superconductivity or formation of various density-wave states. 
The advantage of these systems is the high controllability of model 
parameters with external fields and preparation conditions. 

Many of these systems display low-energy feature that fall outside the standard behavior predicted by
 Landau's Fermi liquid or symmetry breaking theories. In particular, the existence of a non-local (string) order parameter
is proven, both analytically \cite{Shelton1996} and numerically \cite{Schollwock1996},
 to be a characteristic feature of several classes of one-dimensional (1D) and quasi-1D systems. 
\cite{GoNeTs,GiamarchiBook} Such non-local order parameters are topologically protected against any 
local perturbations. The nature of these order parameters and the connection to topological
invariants \cite{GoNeTs} is well understood \cite{Affleck1989} for spin chains with $S\ge 1$. 
For instance, the Haldane conjecture \cite{Haldane83} proposed more than 20 years ago states
that the properties of $SU(2)$ symmetric antiferromagnetic (AF) spin-$S$
Heisenberg chains differ for integer and half-integer spins. The
excitations in the AF Heisenberg chains with half-integer spins are
gapless~\cite{Affleck86} whereas in the integer spin case, a gap is present. The
pure one-dimensional (1D) AF spin-$1/2$ Heisenberg chain can be mapped onto a
Luttinger liquid which allows an exact bosonization treatment, resulting in a well understood gapless
phase. \cite{Affleck86} In contrast, for the AF spin-$1$ Heisenberg chain it is
widely accepted that the excitations exhibit a gap, thanks to extensive numerical~\cite{Nightingale86,Takahashi89,White1993,Deisz93,Golinelli1994,Todo01} and experimental~\cite{Honda1998,Zheludev2003} analysis.

The present understanding of systems of coupled identical $S=1/2$ chains (spin ladders) 
is based on its similarities to larger-spin chains. This similarity allows the one-to-one 
translation of Haldane's conjecture, originally formulated for large-spins chain, onto spin-1/2 ladders with 
an odd or even number of identical legs. The usual assumption about the equivalence of the individual chains constituting the ladder,
is referred below as a symmetric ladder situation.  While the behavior of symmetric ladders has been 
thoroughly investigated, both theoretically and experimentally
\cite{Dagotto1996,GoNeTs}, the case of spin ladders with inequivalent legs (asymmetric ladders) is less understood. 
The behaviour of the spin gap, in particular for the case of a single chain coupled to
 nearly free spins (``dangling spins'')  was recently discussed in
  \cite{Kiselev05b, Kiselev05a,Essler2007,Brunger2008}, but no firm conclusions about the gap scaling in the 
  weak coupling regime were made there. 

\begin{figure}
\includegraphics[scale=0.2]{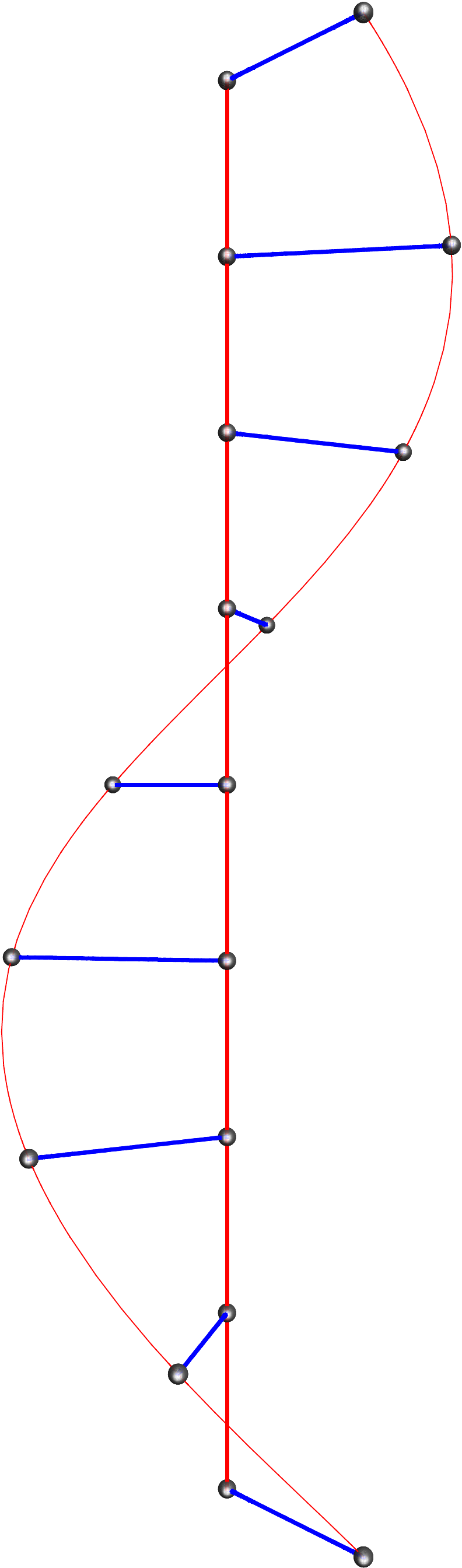}\qquad
\includegraphics[scale=0.6]{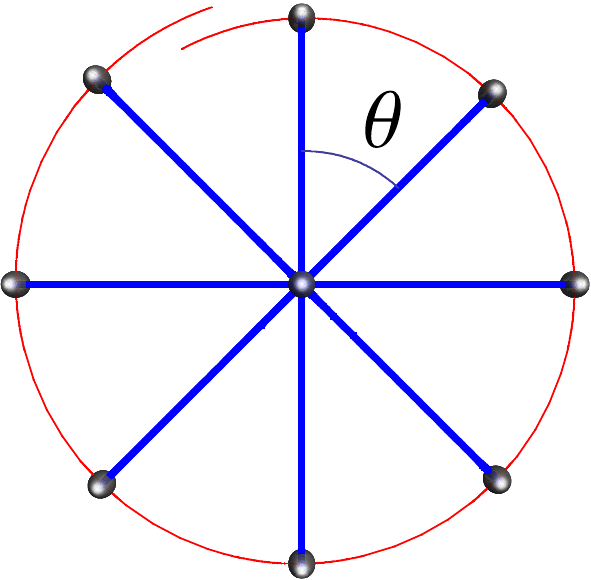}
\caption{\label{fig:SSHL-sketch} (Color online)
         (a) Sketch of the Spiral Staircase Heisenberg Ladder.
         (b) View of the system from the top.
         For $\theta=0$ the model corresponds to the standard
             antiferromagnetically coupled spin-$1/2$ Heisenberg ladder with
             FM rung coupling.
         The case $\theta=\pi$ corresponds to the 1D $SU(2)$ symmetric single-pole ladder
          model.}
\end{figure}

In this work, we consider the Spiral Staircase Heisenberg ladder (SSHL), consisting of 
of two unequal antiferromagnetically coupled spin-$1/2$ chains with
a ferromagnetic (FM) rung coupling $J_{\bot}$. Geometrically, this model may
be understood as a continuous twist deformation of an isotropic 2-leg ladder
with interleg coupling $J_{\|}$ along leg $1$ by an angle $\theta$
(see Fig.~\ref{fig:SSHL-sketch}). As a result of the deformation, the coupling between neighboring sites along
leg 2 is rescaled to the form $J_\| \cos^2\left(\theta/2\right)$ leading to the
Hamiltonian (\ref{eq:SSHL-hamilton1}) below. 

%
As is known from the seminal bosonization work of Shelton, Nersesyan and
Tsvelik, \cite{Shelton1996}  the FM coupling of AF spin chains induces a
Haldane gap proportional to $J_{\perp}$ for weak interleg couplings. However, when the spin velocity on one of the legs vanishes the
bosonization method fails as seen from the fact that a simple formulation of the continuum limit,
on which the bosonization approach relies, is inhibited. Alternative approaches, such as using other analytical or numerical methods, are then demanded.
Beside the theoretical interest, an experimental motivation comes from the fact that the single-pole ladder model at $\theta=\pi$ can been used for modeling a stable organic biradical crystal PNNNO. \cite{Hosokoshi99} 


In our previous work (Ref.~\onlinecite{Brunger2008}), 
we used the quantum Monte Carlo simulations for investigation of asymmetric ladders.
We demonstrated a non-zero value of string order parameter for the whole family of such 
ladders, which confirmed that the system is  in a Haldane phase.  We also presented numerical evidence 
for the smaller energy scale,  $J_\perp^2/J_\|$, associated with Suhl-Nakamura interaction (see below). 
Numerical results for the spin gap were also judged compatible as vanishing as $J_\perp^2/J_\|$, even
though a faster decay could not be ruled out.
These results were consistent with the flow equation calculations of Essler and 
coworkers  \cite{Essler2007}.

In the present paper we show, that the spin gap changes its behavior at small $J_{\perp} \ll J_{\|}$ from
$\Delta \sim J_{\perp}$ in the symmetrical ladder case to $\Delta \sim J_{\perp} \exp( -J_{\|}/J_{\perp})$
 in the single-pole ladder  case. Alternatively, we may think about rungs with a fixed exchange coupling $J_{\perp}$, 
 while the leg exchange increases gradually. We observe that the small-$J_{\|}$ behavior $\Delta \sim J_{\|}$ is the 
 same for all asymmetric ladders, but that important differences appear in the limit $J_{\|} \to \infty$. 
 The symmetrical ladder 
 displays a saturation of the spin gap,  $\Delta \sim J_{\perp}$, while the gap in the single-pole ladder reaches
  a maximum at $J_{\|} \sim J_{\perp}$ and exhibits exponential 
 suppression beyond that scale, according to the above formula.  

The plan of the paper is as follows. We introduce the model and important definitions in Section \ref{sec:setup}, 
where we also provide simple qualitative 
considerations.  Particularly, we explain here the importance of Suhl-Nakamura (SN) indirect 
exchange between dangling spins in the single-pole situation.

We propose analytical approaches to our model in Section \ref{sec:analytic}.
In Section \ref{sec:blocks} we develop a theory, incorporating spin-wave analysis, 
SN interaction and Kadanoff's decimation procedure, which satisfactorily describes the whole body of 
numerical data presented below.  
We stress that several quantities are extracted from the 
numerical data and plotted specifically 
for comparison with the predictions of this effective theory.   
The unusual slow saturation of the spin gap value at large $J_{\perp}$ is discussed in 
Section \ref{sec:app2}.

In Section \ref{sec:numeric} we describe the results of our extensive numerical investigations 
of the problem. Section \ref{sec:exactdiag} discusses  exact diagonalization (ED) results. The appearance
of a SN energy scale $ J_\bot^2/J_\| $ is shown here. Given the long-range character 
of the SN interaction which require large systems to be appreciated, we resort to large scale 
quantum Monte Carlo (QMC) simulations in the next subsection.  
With this approach, we investigate the form of the spin correlation functions
in Section \ref{sec:qmc}, compute the spin gap in Section \ref{sec:QMCgap} as
well as the string order parameter in Section \ref{sec:SOP}, characteristic of the Haldane phase.
The most challenging case of the single-pole ladder gives rise to the largest uncertainties for the spin gap, 
and we focus on this system with the use of the DMRG method, as described in Section \ref{sec:DMRG}. The DMRG 
results do not give a principal advantage over QMC findings, but they provide a strong independent verification of the 
observed form of the spin gap, namely an exponential suppression 
at small $J_{\perp}$. 

In Appendix \ref{sec:meanfield}, we analyze the qualitative changes in the spectra
of symmetric and single-pole ladders within a Jordan-Wigner mean-field calculation.  
Technicalities of spin-wave theory for long-range interaction are given in Appendix \ref{sec:app1}.
We finally present our conclusions in Section \ref{Sec:Conclusions}. 

\section{problem setup and qualitative considerations} 
\label{sec:setup}

We study the low-energy physics of the SSHL system, characterized by the 
following Hamiltonian:
\begin{eqnarray}
 {\mathcal{H}}
&=& J_\| \sum_i\left(
     {\mathbf{S}}_{1,i}\cdot  {\mathbf{S}}_{1,i+1}+
    \cos^{2}\left(\tfrac{\theta}{2}\right)
     {\mathbf{S}}_{2,i}\cdot  {\mathbf{S}}_{2,i+1}\right)
    \nonumber\\
&-& J_\perp\sum_{i}  {\mathbf{S}}_{1,i}\cdot  {\mathbf{S}}_{2,i}\,\, .
\label{eq:SSHL-hamilton1}
\end{eqnarray}
Here, $ {\mathbf{S}}_{\alpha,i}$ is a spin-$1/2$ operator acting on leg
$\alpha$ and lattice site $i$ and  $J_\|>0$ sets the energy scale. The single-pole ladder model \cite{Kiselev05b,Aristov2007a, Brunger2008} corresponds to $\theta=\pi$.

It is convenient to reformulate the model~Eq.(\ref{eq:SSHL-hamilton1}) in terms of
new variables,
\begin{eqnarray}
 {\mathbf{S}}_{i}= {\mathbf{S}}_{1,i}+ {\mathbf{S}}_{2,i}
\,\, ,\quad
 {\mathbf{R}}_{i}= {\mathbf{S}}_{1,i}- {\mathbf{S}}_{2,i}\,\, ,
\label{eq:rungoperators}
\end{eqnarray}
defined on the rung. The Hamiltonian then reads
\begin{eqnarray}
 {\mathcal{H}}
&=& \frac{J_{\|}}{4}\sum_{i}
    \left(1+\cos^{2}\left(\tfrac{\theta}{2}\right)\right)
    \left( {\mathbf{S}}_{i}\cdot {\mathbf{S}}_{i+1}+
     {\mathbf{R}}_{i}\cdot {\mathbf{R}}_{i+1}\right)
    \nonumber\\
&+& \frac{J_{\|}}{4}\sum_{i}
    \sin^{2}\left(\tfrac{\theta}{2}\right)
    \left( {\mathbf{S}}_{i}\cdot {\mathbf{R}}_{i+1}+
     {\mathbf{R}}_{i}\cdot {\mathbf{S}}_{i+1}\right)
    \nonumber\\
&-& \frac{J_{\bot}}{4}\sum_{i}
    \left( {\mathbf{S}}^{2}_{i}- {\mathbf{R}}^{2}_{i}\right)\,\, .
\label{eq:SSHL-hamilton2}
\end{eqnarray}
The set of operators $ {\mathbf{S}}_{i}$ and $ {\mathbf{R}}_{i}$ fully
defines the $o_{4}$ algebra. \cite{Kiselev05a,Kiselev05b}

Let us define the retarded spin response function in the ladder situation
 \begin{equation}
 \begin{aligned}
\chi_{jl}( q,\omega)&= - i \int_0^\infty dt \sum_{n}
e^{i\omega t - i q n}    \langle[ {S}^{z}_{j,1}(t) , {S}^{z}_{l,1+n}] \rangle , 
\\ &= \int d \omega'   \frac{S_{jl}(q,\omega')\,  } { \omega- \omega'+\ii0}
\end{aligned}
\end{equation} 
with $j,l = 1, 2$. 
At zero temperature the dynamic structure factor (spectral weight), given by 
$S_{jl} ({q},\omega) = -\pi^{-1} {\rm Im}\,  \chi_{jl}( q,\omega)$ is represented as 
\begin{equation}
\begin{aligned}
S_{jl}({q},\omega) &=  \sum_{n}\langle 0| {S}^{z}_{l}(q) |n \rangle 
\langle n| {S}^{z}_{j}(-q) |{0} \rangle  \delta(E_{n}-E_{0}-\omega)
\end{aligned}
\label{eq:dynspinfactor}
\end{equation}
 where  
$|{0}\rangle$ stands for the ground state with the energy $E_{0}$, 
 the sum runs over all eigenstates $|n\rangle$ of the Hamiltonian
with energies $E_{n}$ and ${S}^{z}_{l}(q)$ is a Fourier transform of ${S}^{z}_{ln}$.
We also define the symmetrized combinations, 
  \begin{equation}
  \begin{aligned} 
  S({q},\omega)
&= S_{11}({q},\omega) + S_{22}({q},\omega) + S_{12}({q},\omega)+
S_{21}({q},\omega),
 \\
   R({q},\omega)
&=   S_{11}({q},\omega) + S_{22}({q},\omega)  -  S_{12}({q},\omega) - 
S_{21}({q},\omega)  ,
    \end{aligned}
\end{equation}
which are response functions for operators $S^z$ and $R^z$ in (\ref{eq:rungoperators}), respectively. 

Below we also use imaginary time correlation functions
  \begin{equation}
  \begin{aligned}
\langle  {S}^{z}_{q}(\tau) {S}^{z}_{-q}(0)\rangle
&= \int d\omega\,
    \frac{e^{-\tau\omega}}{1-e^{-\beta\omega}} S(q,\omega)
     \\
\langle  {R}^{z}_{q}(\tau) {R}^{z}_{-q}(0)\rangle
&= \int d\omega\,
    \frac{e^{-\tau\omega}}{1-e^{-\beta\omega} } R(q,\omega)
    \end{aligned}
    \label{corr-imagtime}
\end{equation}
 with $\beta =  1/T$. 

Let us now qualitatively discuss the situation.  In the strong coupling region,  $J_{\bot}/J_{\|} \rightarrow \infty $,  and for all
possible twist angles $\theta$  triplets on the rungs become more and more favorable 
such that the Hamiltonian of Eq.~(\ref{eq:SSHL-hamilton2})  reduces to a pure spin-1
{\it effective} Hamiltonian:
\begin{eqnarray}
 {\mathcal{H}}_{\mathrm{eff}}
&=&J_{\mathrm{eff}}\sum_{i}
  {\mathbf{S}}_{i}\cdot {\mathbf{S}}_{i+1}
\label{eq:Jeff}\\
J_{\mathrm{eff}}
&=&\tfrac{J_{\|}}{4}\left(1+\cos^{2}(\theta/2)\right)\,\, .
  \nonumber
\end{eqnarray}

In units of effective coupling  
$J_{\mathrm{eff}}$ (and thus irrespective of the twist angle) the spin gap of our  model
scales to the Haldane gap $\Delta_{H}/J_{\mathrm{eff}}=0.41048(6)$. \cite{Todo01}

In the weak coupling region the situation is more delicate and
depends  on the twist $\theta$.  Let us discuss here the limiting cases. 

For the symmetric ladder, $\theta=0$, the gap opens as $\Delta \sim J_{\perp}$, as obtained by 
bosonization and also qualitatively reproduced in the simplified mean-field picture below. 
For the fully asymmetric single-pole ladder $\theta=\pi$ the mean-field calculation predicts a 
sub-linear behavior of the gap $\Delta \sim J_{\perp}^2/J_{\|}$.  The bosonization treatment becomes 
problematic in this case, as we cannot apply the continuum approach to the chain of dangling spins attached 
to the main leg. The assumption of the finite Fermi (sound) velocity in the subsystem breaks 
down and  $J_{\perp}$ cannot be used as a perturbation in the implicitly assumed hierarchy of scales 
$J_{\perp} \ll J_{\|} \cos^{2}(\theta/2) < J_{\|}$. 

Instead, we should start with the picture of a degenerate band of dangling spins, whose degeneracy is lifted 
by indirect exchange between these spins through the main leg. 
This phenomenon is known as 
a Suhl-Nakamura interaction \cite{Suhl57,Nakamura58,Aristov94} for the case of a dilute system of extra 
spins in a magnetic host. It has a direct analogy with the RKKY interaction where itinerant electrons mediate 
a long range spin-spin interactions between  localized  spins
(see Ref.~\onlinecite{Aristov1997a} and references therein).
In second order perturbation theory the SN interaction reads: 
\begin{eqnarray}
J_{SN}\propto J^{2}_{\perp}\chi( {q},\omega=0)\,\, ,
\end{eqnarray} 
where $\chi( {q},\omega=0)$ is the spin susceptibility of the spin-$1/2$
Heisenberg chain.
We thus arrive at a similar energy $J_{\perp}^2/J_{\|}$ but now it stands for the SN-induced 
bandwidth, not for the spin gap.

It is important to remark that both the mean-field treatment and the SN energy scale estimate provide us \emph{only} 
 with an upper bound for the gap value, $J_{\perp}^2/J_{\|}$. However this is likely a strong overestimation 
 as quantum fluctuations, which are important in 1D, are not accounted for. 
 In Sec.\  \ref{sec:analytic} we present analytic arguments in favor of  
a gap vanishing faster than with a power law.
In Sec.\  \ref{sec:numeric} we show, that the whole body of numerical data supports 
this scenario.


\section{Analytic approaches and interpretation} 
\label{sec:analytic}

In most cases, it is instructive to map the system of spins onto a system of 1D spinless fermions. 
We show in Appendix \ref{sec:meanfield}, that such Jordan-Wigner transformation and a subsequent
mean-field theory analysis predict qualitative changes in the behavior of the gap with $\theta$. 
Particularly, Eq.\  (\ref{gapMFT}) there indicates
that the prefactor in the linear law, $\Delta \sim J_{\perp}$, diminishes with decreasing $\cos\tfrac\theta2$, 
and that in the extreme case 
of single-pole ladder  ($\cos \tfrac\theta2 \to 0$), the gap vanishes faster than $|J_\perp|$. 
These observations 
are qualitatively confirmed by our numerical simulations. At the same time, the attempt to compare 
the mean-field results for the fermionic theory 
with the gap extracted from QMC data shows that the mean-field theory overestimates the gap by an
 order of magnitude.  

In view of this fact, we perform a different type of analysis in the next subsections. 

\subsection{Decimated blocks and effective spins}
\label{sec:blocks}

Below we propose a scenario, which assumes different behavior of spin dynamics at high and low energies, as
separated by the scale $J_{\perp}$. This scenario is explicitly formulated for the single-pole ladder, which represents
the most intriguing and difficult case, as seen in the numerics. The extension of this scenario to $\theta \neq \pi$ is 
also briefly 
discussed at the end of this subsection. The proposed theory allows to semi-quantitatively explain all features 
observed in the large-scale QMC and DMRG studies. 

For the high energies in consideration $\epsilon > J_\perp$, the dangling spins should be considered as freely attached 
to the main chain. These energies correspond to short times, 
$t< J_\perp^{-1}$, and distances, $x < J_{\|}/J_{\perp}$. 
 Alternatively, one may think in terms of a higher temperature $T>J_\perp$, which smears all fine features of the 
 spectrum and  
leads to exponential decay of correlations beyond the temperature correlation length $\sim J_{\|}/T$. 
The inverse length scale along the leg, $\xi^{-1}$, associated with the crossover to the low-energy dynamics is defined 
by the relation $J_\perp \sim \xi^{-1} J_\|$, showing that soft spinon excitations in the HAF spin-1/2 chain, with momenta
  $q\alt \xi^{-1}$ are strongly intertwined with triplet-singlet transitions on the rungs.

At these shorter time scales the classification in terms of singlet and triplet on the rung is not very appropriate. 
It means that the dangling spins are viewed as almost decoupled from the main chain: the situation is best described in terms of dangling spins coupled to each other by the Suhl-Nakamura interaction. The long-range character of SN interaction leads to an almost flat dispersion of the magnon excitations in  
the subsystem of dangling spins, the estimated SN energy scale not exceeding the crossover energy $J_\perp$.  

At smaller energies, $\epsilon < J_\perp$, a picture of already formed 
triplets on the rungs is more appropriate. The discussion of 
the dynamics reduces to the rotations of the effective spins $S=1$. 
Further, these rung triplets are interlaced by the leg interaction into large effective spin blocks of size $\xi$. 
Despite a possibly large spatial size of the block, the AF character of the main leg interaction selects the smallest possible total spin state of this block. Discarding the non-magnetic singlet, we focus on 
the block triplet state, i.e., when $\xi$ spins $S=1$ are combined into a new effective spin 
$S=1$ of the block. As a result, we have a model with nearest-neighbors interaction between large blocks. Such a model 
should display a Haldane gap, whose value can be determined from usual arguments. 
 
Let us start with the shorter distances and higher energies. In this case the SN interaction between dangling spins $S_{2,i}$ and $S_{2,i+x}$ has the form $V(x) S_{2,i} S_{2,i+x}$ with 

\begin{equation}
V(x) = J_\perp^2 \int \frac{dq}{2\pi} \chi(q,\omega=0) e^{iqx} 
\sim J_\perp^2/J_\| (-1)^x \ln(\xi/x)
\label{SNint}
\end{equation}
with $\chi(q,\omega)$ is the response function for the HAF $S=1/2$ 
model. \cite{GoNeTs} For our purposes it suffices to approximate 
$\chi(q,\omega=0)$ by $J_\|^{-1}|\cos (q/2)|^{-1}$. The $1/q$ 
singularity  
near $q=\pi$ shows that $V(x)$ is sign-reversal and logarithmically 
decaying with distance,  $V(x) \sim (-1)^x \ln(\xi/x)$. The scale 
$\xi$ in the argument of the logarithm,appears 
here as a parameter which will be further determined by a self-consistency criterion. 
Technically it is assumed that the allowed energies, $\omega_q$, of the spin-wave continuum are restricted from below,  $\omega_q \agt J_\|/\xi$.

A chain of dangling spins, coupled by long-range SN interactions  (\ref{SNint}), behaves differently from the standard HAF 
model with nearest-neighbor interaction. A simple spin-wave analysis is then indispensable here. Such an analysis cannot give a 
correct form of correlation functions, but delivers a qualitative information about the spectrum. 
\cite{Affleck1999,  Aristov2004}

Using the formulas listed in Appendix \ref{sec:app1} we conclude that 
the spin-wave spectrum in the system of dangling spins is given by the expression $\omega_{k} = \sqrt{g_k g_{k+\pi}}$,
 with $g_k = V(\pi) - V(\pi + k)$. Approximating the range function $V(\pi + k) \simeq (J_\perp^2/J_\|) (\sin^2(k/2) + 
 \xi^{-2})^{-1/2} $  we have 
\begin{equation}
\begin{aligned}
\omega_{k+\pi} \simeq &
(J_\perp^2/J_\|)  \xi 
\sqrt{ 1 - (1 + \xi^{2}\sin^2(k/2) )^{-1/2} }
 \\  
 \simeq &
 (J_\perp^2/J_\|) \xi, 
\quad |k|\gg \xi^{-1} 
 \\ \simeq &
(J_\perp^2/J_\|) \xi^2 |k|, 
\quad |k|\ll \xi^{-1}  
\end{aligned}
\label{SN-magnons}
\end{equation}

Next we make the natural assumption that the top of the SN-induced band coincides (at least by the order of magnitude) with the logarithmic low energy cutoff introduced above, leading to $(J_\perp^2/J_\|) \xi \sim
J_\| /\xi$ or  
\begin{equation}
\xi \sim  J_\| / J_\perp
\end{equation} 

Remarkably, this estimate shows that the low-energy spin-wave dispersion (\ref{SN-magnons}) is characterized 
by the same velocity $\omega_{k+\pi} \sim J_\| |k|$ as the higher-energy excitations in the main leg. 
This is despite the fact that, strictly speaking, the low energies $\epsilon \ll J_\perp$ should be considered with a different approach as proposed below.  
Notice that the described spectrum resembles a simple picture of hybridization between the linear spectrum and 
initially degenerate band at non-zero energy
$\sim J_{\perp}$, with the crossing level repulsion phenomenon.   

At small energies $\epsilon \ll J_\perp $, we expect that the picture of spinons (in the main leg) scattering on the dangling spins, becomes inadequate. 
The dynamics of spins on the rung is characterized in terms of soft triplet dynamics, while the transitions to singlet 
state with the (now) large energy $J_\perp$ are discarded. On the same ground, we discard spin-wave 
continuum excitations with energies $\epsilon \agt J_\perp \sim J/\xi $, i.e.,  the description now has changed from the individual sites along the leg to entire blocks of length $\xi$. We may thus think in terms of the effective spin $S=1$ on the rung, and these individual spins $S=1$ are 
antiferromagnetically coupled to each other in the large block. 

Furthermore, we can characterize the whole block $\xi$ of spins 1 by its lowest non-trivial multiplet, 
a spin 1 again. However now it is an object defined at a much larger spatial scale. The trivial low-energy 
multiplet in such $\xi$-block is a spin-singlet state, which obviously drops out from the soft spin dynamics. 

The typical energy spacing between the resulting spin-triplet of the $\xi$-block and the higher multiplets 
is estimated again as $J_\|/\xi \sim J_\perp$. We expect that this lowest triplet state is non-degenerate, i.e. other 
triplets are higher in energy. Below we refer to this lowest triplet state as $\xi$-triplet. 

We note in passing that the estimate $\xi \sim J_{\|}/J_{\perp}$ follows also from the argument 
presented in Ref.~\onlinecite{Essler2007}. It was suggested there that the spinons, propagating along the main leg by distance $m$ and characterized 
by a typical energy $J_{\|}$, break the pre-formed rung triplets resulting in an energy cost $\sim m J_{\perp}$. 
The resulting confining potential should, in principle, restrict the motion of spinons to distances $\sim J_{\|}/J_{\perp}$. 

Our way of constructing $\xi$-triplet resembles Kadanoff's decimation procedure in the description of critical 
phenomena. \cite{Huang1987}
The new lattice of large blocks contains spins 1, which are denoted below by $S_{\xi,n}$ (here $n$ numbers 
a position in a new lattice) and are coupled to each other by a nearest-neighbor interaction. In fact, only the edge 
spins of each $\xi$ block are responsible for this interaction. Adopting the notation that the weight of these edge 
spins in the $\xi$-triplet is  $w_{1} \ll 1$ at $\xi\gg1 $ (see below), we write 
explicitly for odd $\xi$ 
\begin{equation}
S_{1,j}^\alpha =  w_{1} S_{\xi,n}^\alpha (-1)^{j+n}
\end{equation}
where $ n = \lfloor j/\xi \rfloor$ and   $\lfloor \ldots \rfloor$ stands for the floor function. 
Similarly, $S_{2,j}^\alpha =  w_{2} S_{\xi,n}^\alpha (-1)^{j+n}$. 
The phase $(-1)^{j }$ accounts for the AF 
character of the contributing spins in the $\xi$-block and the additional phase shift $(-1)^{n}$ is introduced to restore 
the translational invariance in the blocks picture. 

The definition of the above weight $w_{j}$ is as follows. Assume that the lowest multiplet in the $\xi$-block is a triplet 
$|T,m \rangle$, spanned by the spin-1 operators $S_{\xi,n}^\alpha$. Then, up to a sign, 
the edge operators of the block act as 
\begin{equation}
 \langle T,m' |S_{1,j}^\alpha |T,m \rangle =
 w_{1} \langle T,m' | S_{\xi,n}^\alpha  |T,m \rangle , 
\end{equation} 
and similarly for the weight $w_{2}$ of $S_{2,j}^\alpha$. The block Hamiltonian on the decimated lattice reads
\begin{equation}
\sum _n  J_\| S_{1,n\xi} S_{1,n\xi +1 } \to 
\sum _n w_{1}^2 J_\| S_{\xi,n} S_{\xi,n+1 }
\end{equation}
with $n = 1,\ldots L/\xi$. 
This model should exhibit a Haldane gap, but the value of this gap is strongly diminished. 

Employing a standard albeit simplified approach used in the original Haldane paper,  
 we apply the linearized spin-wave theory outlined in Appendix \ref{sec:app1} 
 and obtain the magnon dispersion, 
$\omega_k = 4 w_{1}^2 J_\| \sin (k)$,  
where $k$ is the wave vector on the decimated lattice $k = 2\pi n (\xi/L) $ with $n=1,2,\ldots$. 
It implies that the velocity of the lowest-lying excitations with respect to the real lattice
is given by $\tilde v =  4 w_{1}^2 \xi J_\|$. With the plausible assumption that this 
low-energy estimate coincides with the high-energy one, we obtain $\tilde v \sim J_\|$ and hence $w_{1}^2 \sim \xi^{-1}$. 
Notice that it corresponds to a bandwidth $4 w_{1}^2 J_\| \sim J_{\perp}$, in accordance with  Eq.\ (\ref{SN-magnons}).

We require that the zero-point magnon fluctuations cancel the local magnetization in the usual formula 
$S_{1,j}^z = 1/2 - S_{1,j}^- S_{1,j}^+$, i.e., we write $\langle S_{1,j}^- S_{1,j}^+ \rangle = 1/2$. 
For the relevant low-lying modes, we have $S_{1,j}^- = w_{1} \sqrt{2} a_n^\dagger$, where  
 $a_n^\dagger$ is magnon creation operator in the $n$th $\xi$-triplet. 
We then obtain the relation 

\[
1/2 \simeq 2w_{1}^2 
\int_{q_0}^{\pi/2} \frac{dk}{\pi} \left(\frac{1}{\sin k }
- 1 \right)
\]
where we introduced the cutoff wave-vector $q_0$, 
related to the Haldane gap $\Delta= 4 w_{1}^2 J_\| \sin q_0$. This leads to the estimate $ q_0 \sim   \exp (-\pi / 4 w_{1}^2 )$ 
and 

\begin{equation}
\Delta \sim w_{1}^2 J_\|  \exp\left (-\frac {\pi}{4w_{1}^2 } \right ) 
\end{equation}

Let us discuss the decay of correlations at the distances $r \gg\xi $ ; we take $r/\xi = x $ integer for simplicity. We have 
\begin{eqnarray}
\langle S_{1,j} S_{1,j+r}\rangle &\simeq &
(-1)^{r+x}
w_{1}^2  \langle S_{\xi, n} S_{\xi,n+x}\rangle   \nonumber \\
&=& (-1)^{r} w_{1}^2  
\int_{-\pi}^{\pi } \frac{dk}{4\pi} 
\frac{1-\cos k}{\sqrt{q_0^2 +\sin^2 k} }
e^{i(\pi +k) x} \nonumber 
\\
&\simeq &
(-1)^{r} \frac{w_{1}^2}{\pi} K_0(r q_0 /\xi)  
\end{eqnarray}

The above assumption that $w_{1}^2 \sim \xi^{-1}$ leads to the following final formulas    
\begin{equation}
\begin{aligned}
\Delta &\sim  J_{\perp}  \exp\left (-\frac{J_{\|}}{J_{\perp}} c_{2} \right ) \\
 \langle S_{1,j} S_{1,j+r}\rangle &\sim 
 (-1)^{r} \frac{J_{\perp}}{\pi J_{\|}} K_0(r \Delta/ J_{\|})  
\end{aligned}
\label{eq:final}
\end{equation}
with $c_{2} \sim 1$.  
Our scenario also suggests that the long-range behavior of correlations (\ref{eq:final}) takes place for both main leg 
and dangling spins at distances $r \agt \xi \sim J_{\|}/J_{\perp}$. 

Comparing the above formulas (\ref{eq:final}) to our numerical findings below, see  
Fig.\  \ref{fig:QMCcorrelations}  and Fig.\ \ref{Fig:Scales}, 
 we verify that, indeed,  $w_{j}^2 \propto J_\perp$.  

Finally, let us briefly comment on the situation where there is a weak exchange along the second leg, $\theta \simeq \pi$ 
and $J_{2} = J_{\|} \cos^{2} \tfrac \theta2 \ll J_{\|}$. Repeating the steps of the analysis leading to Eq.\ (\ref{SN-magnons}), 
we observe that the SN induced band appears on the background of usual exchange. Roughly, one can write 
$V(k) \sim -J_{2} \cos k  + J^{2}_{\perp}/J_{\|} |\cos(k/2)|^{-1} $ and the bandwidth is estimated 
as $J_{2} + \xi J^{2}_{\perp}/J_{\|} $. This scale should not exceed the  low-energy cutoff for the otherwise 
long-range SN interaction,  $J_{\|}/\xi$, which leads to the corrected estimate, $\xi^{-1} \agt \max [ J_\perp/J_\| , \cos^{2} 
\tfrac \theta2] $. The above scenario remains valid, as long as $\xi^{-1} \ll 1$, which imposes restrictions on  
$\theta$. We see in Fig.\ \ref{fig:QMCgap}a, that the gap at $\theta = 8\pi/9$ ($\cos^{2} \tfrac \theta2 \simeq 0.03 $)
behaves qualitatively similar to the one at $\theta =\pi$, whereas the behavior 
of the gap at $\theta = 7\pi/9$ ($\cos^{2} \tfrac \theta2 \simeq 0.12 $) is apparently different and closer 
to the one of the symmetric ladder, $\theta =0$. For our semi-quantitative level of consideration these 
conclusions appear consistent and satisfactory.  It was implied in Ref.\ \cite{Brunger2008} that 
the scaling  of the gap with $J_{\perp}$ changes at a critical $\theta_{c} \sim \pi$. The above consideration 
suggests that there is no critical $\theta$ but rather a smooth crossover between two regimes, cf. also the estimate 
$\theta \simeq 0.54 \pi$ in the next subsection for the crossover in strong rung coupling behavior.

\subsection{Strong rung coupling limit}
\label{sec:app2}

Let us consider the limit of strong rung coupling, $|J_\perp| \gg
J_\|$. In the zeroth order of the small parameter $J_\|/|J_\perp|$, we
have the effective Hamiltonian, (\ref{eq:Jeff}). The
perturbation is given by the operators  $R^\alpha_i$ in (\ref{eq:SSHL-hamilton2}),
which connect the spin-1 rung sector, spanned by operators $S^\alpha_i$,
to the high-energy singlet rung state.
The perturbing part is given by

\begin{equation}
\begin{aligned}
\hat{\mathcal{H}}_{\mathrm{int}}
&=
 \sum _{i\alpha} \left(J_{\textsf{RQ}}  R^\alpha_i Q^\alpha_{i}
 +   J_{ \textsf{RR}}  R^\alpha_i R^\alpha_{i+1} \right)
 \\
  Q_i^\alpha &= (S^\alpha_{i-1} + S^\alpha_{i+1}) \\
    J_{\textsf{RQ}} &= \tfrac14 {J_\|} {}  (1 - \cos^2 (\tfrac\theta2)) , 
\end{aligned}
\end{equation}
and $   J_{\textsf{RR}} = \frac {J_\|}4 (1 + \cos^2 (\theta/2) = J_{\mathrm{eff}}$. 
The leading correction to the effective Hamiltonian (\ref{eq:Jeff}) is
obtained in second order of perturbation in
$\hat{\mathcal{H}}_{\mathrm{int}}$, by considering the virtual
transitions to singlet states separated by the energy $|J_\perp|$.
This derivation is
similar to the one of the $t$-$J$ model from the large-$U$ Hubbard
model. We have:
\begin{equation}
\begin{aligned}
\hat{\mathcal{H}}_{\mathrm{eff}}^{(1)}
=&
\hat{\mathcal{H}}_{\mathrm{eff}}^{(1A)} +
\hat{\mathcal{H}}_{\mathrm{eff}}^{(1B)}
\\
\hat{\mathcal{H}}_{\mathrm{eff}}^{(1A)}
=& -\frac{  J_{\textsf{RQ}}^2}{|J_\perp|}
\sum_{i\alpha\beta}
R_i^\alpha R_i^\beta Q_i^\alpha Q_i^\beta
\\
\hat{\mathcal{H}}_{\mathrm{eff}}^{(1B)}
=& - \frac{  J_{\textsf{RR}}^2}{2|J_\perp|}
\sum_{i\alpha\beta}
R_i^\alpha R_i^\beta
R_{i+1}^\alpha R_{i+1}^\beta
\end{aligned}
\end{equation}
Using the identity
\begin{equation}
R_j^\alpha  R_j^\beta + S_j^\alpha  S_j^\beta  =
\delta_{\alpha\beta} + i \epsilon_{\alpha\beta\gamma} S_j^\gamma
\end{equation}
one can eventually arrive to
\begin{equation}
\begin{aligned}
\hat{\mathcal{H}}_{\mathrm{eff}}^{(1A)}
=& - \frac{  J_{\textsf{RQ}}^2}{|J_\perp|}
\sum_{i}[ {\bf Q}_i {\bf Q}_i - {\bf S}_i {\bf Q}_i
- ({\bf S}_i {\bf Q}_i)^2]
\\  =& 
 2\frac {  J_{\textsf{RQ}}^2}{|J_\perp|}
\sum_{i}\left[
{\bf S}_i {\bf S}_{i+1}+ ({\bf S}_i {\bf S}_{i+1})^2
\right. \\ &\left.
- {\bf S}_{i-1}{\bf S}_{i+1} +
({\bf S}_i {\bf S}_{i-1})({\bf S}_i {\bf S}_{i+1}) \right]
\\
\hat{\mathcal{H}}_{\mathrm{eff}}^{(1B)}
= &- \frac{  J_{\textsf{RR}}^2}{2|J_\perp|}
\sum_{i}({\bf S}_i {\bf S}_{i+1})^2
\end{aligned}
\label{heff1}
\end{equation}
The effective Hamiltonian (\ref{heff1}) is rather complicated, containing quartic combinations of spins, 
and we propose here its qualitative analysis. 
The main term, Eq.\  (\ref{eq:Jeff}), results in antiferromagnetic correlations of 
 the spins $S_{i}$  at adjacent sites,  whereas 
the next-to-nearest neighboring spins are aligned ferromagnetically.  We thus expect  ${\bf Q}_{i}$ to be
 in the state $Q=2$, and ${\bf Q}_i {\bf Q}_i = Q(Q+1) = 6$. Next,   let ${\bf Q}_{i}$  and ${\bf S}_i$ be added into a multiplet 
of total spin $p$, so that $ {\bf S}_i {\bf Q}_i = \tfrac12 ( ({\bf S}_i+{\bf Q}_i)^{2} - {\bf S}_i^{2} - {\bf Q}_i^{2})=
 \tfrac 12 p(p+1) - 4  $. One can check, that 
$[ {\bf Q}_i {\bf Q}_i - {\bf S}_i {\bf Q}_i
- ({\bf S}_i {\bf Q}_i)^2]   = 0, 6, 0$ for states with  $p=1, 2, 3$, respectively. It means that the mostly AF correlated
 state with $p=1$ obtains a higher energy due to $\hat{\mathcal{H}}_{\mathrm{eff}}^{(1A)}$, while the less antiferromagnetically 
 aligned state $p=2$ leads to an energy gain 
 $- 6{  J_{\textsf{RQ}}^2}/{|J_\perp|}$. Alternatively, we may use the second line in the
 representation of $\hat{\mathcal{H}}_{\mathrm{eff}}^{(1A)}$ in (\ref{heff1}) and estimate it as 
 $\simeq 2\left[ x +2 x^{2} - 1   \right]   {  J_{\textsf{RQ}}^2}/{|J_\perp|}$.
 Here $ {\bf S}_i {\bf S}_{i+1}\equiv x  = -2, -1, 1$ for the total spin of the pair,  ${\bf S}_i + {\bf S}_{i+1}$, equalling 
 $0, 1, 2$, respectively;  we also approximate ${\bf S}_{i-1}{\bf S}_{i+1} \simeq 1$. Combining it with 
 $\hat{\mathcal{H}}_{\mathrm{eff}}^{(1B)}$,  we have the estimated energy per unit cell  
\begin{eqnarray}  
\delta E 
&\simeq& 
 \frac{2J_{\mathrm{eff}}^{2}}{|J_\perp|} \left[ 
Y (x +2 x^{2})
 - \frac {x^{2}}4  
  \right]
  \label{heff1-estim}
  \\
Y&=&   \left( \frac{1 - \cos^2 (\theta/2)}{1 + \cos^2 (\theta/2)} \right)^{2} \nonumber
\end{eqnarray} 

In a simplified picture, we can compare the energy difference, $\Delta E_{ts}$, between 
the bond triplet state, $x=-1$, and the bond singlet state, $x=-2$. From  (\ref{heff1-estim}) it follows that 
this energy difference per unit cell is 
\begin{equation}
\Delta E_{ts} =  J_{\mathrm{eff}} + \frac{2J_{\mathrm{eff}}^{2}}{|J_\perp|} 
\left( - 5 Y   +\tfrac {3}4    \right)
\label{eq:Ets}
\end{equation}

For symmetric ladder, $\theta =0$  and $Y=0$, 
the correction  (\ref{heff1-estim}) favors the bond singlet and $\Delta E_{ts}$ is always positive. 
For the single-pole ladder, 
$\theta =\pi$ and $Y=1$, the expression (\ref{eq:Ets}) shows that the corrections 
$\sim J_{\mathrm{eff}}^{2}/ {|J_\perp|}$ are important even for $ J_\perp \sim 10  J_{\mathrm{eff}}$, due 
to a large numerical prefactor.  
The sign of the correction in  (\ref{eq:Ets}) changes at $Y=3/20$ or $\theta \simeq 0.54 \pi$, and indeed we observe 
a slower saturation of the gap at $J_{\perp} \gg J_{\|}$ for $\theta > \pi/2$.

Roughly, we can regard $\Delta E_{ts}$ as a new value of 
$J_{\mathrm{eff}}$  in Eq.\ (\ref{heff1}), and it implies that the gap at $\theta > \pi/2$ should follow the 
law $\Delta \simeq 0.41 J_{\mathrm{eff}} (1- c(\theta) J_{\mathrm{eff}}/{|J_\perp|}) $ with $c(\theta) \sim 1$.  
At large value of  $c(\theta)$, the intermediate region $1\alt J_{\perp}/J_{\|} \alt c(\theta)$ becomes rather extended.  
The detailed description 
of $\Delta (J_{\perp})$ at intermediate $J_{\perp}$  is beyond the scope of this study.

\section{Numerical analysis}
\label{sec:numeric}

%
%
%
\subsection{Exact Diagonalization}
\label{sec:exactdiag}
In this section we analyze the SSHL model by means of exact diagonalization (ED)
methods using the Lanczos algorithm. \cite{Lanczos50,Dagotto94} Even though
ED methods are limited to small systems, they provide considerable insight.
We start our analysis with a study of the spin excitations $S_{jl}({q},\omega)$, Eq.\ (\ref{eq:dynspinfactor})

%
\begin{figure}
 \includegraphics*[width=0.45\textwidth]{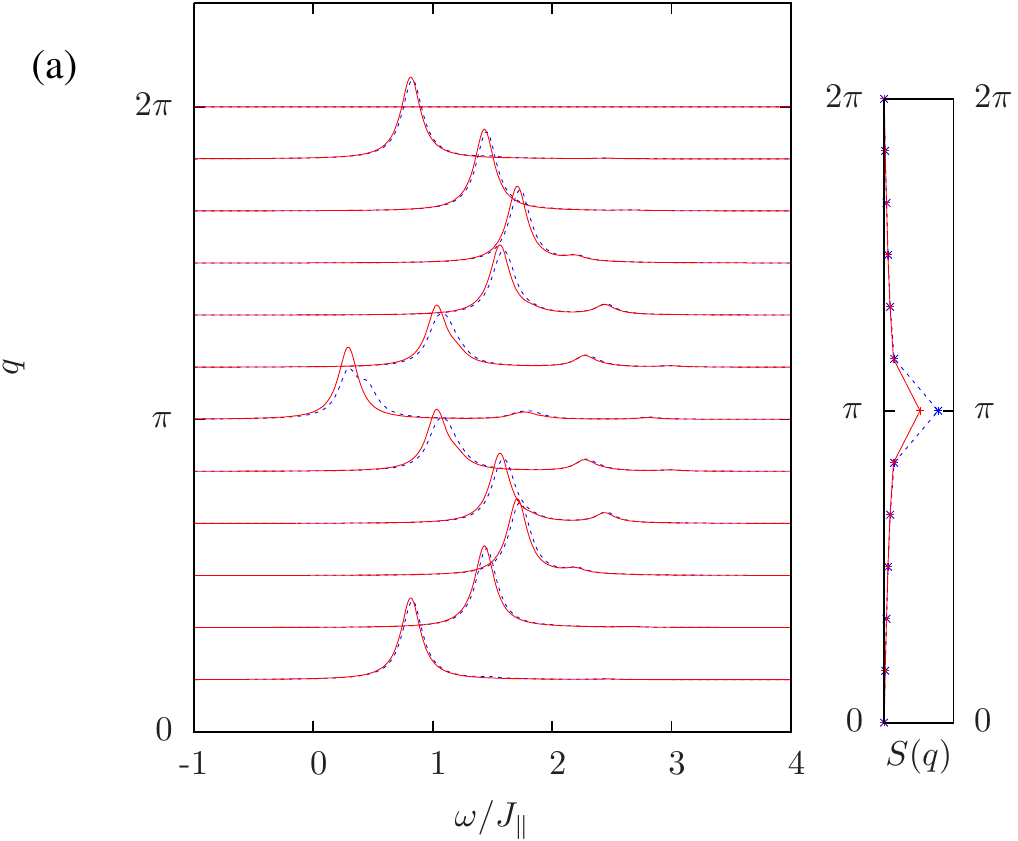}
 \includegraphics*[width=0.45\textwidth]{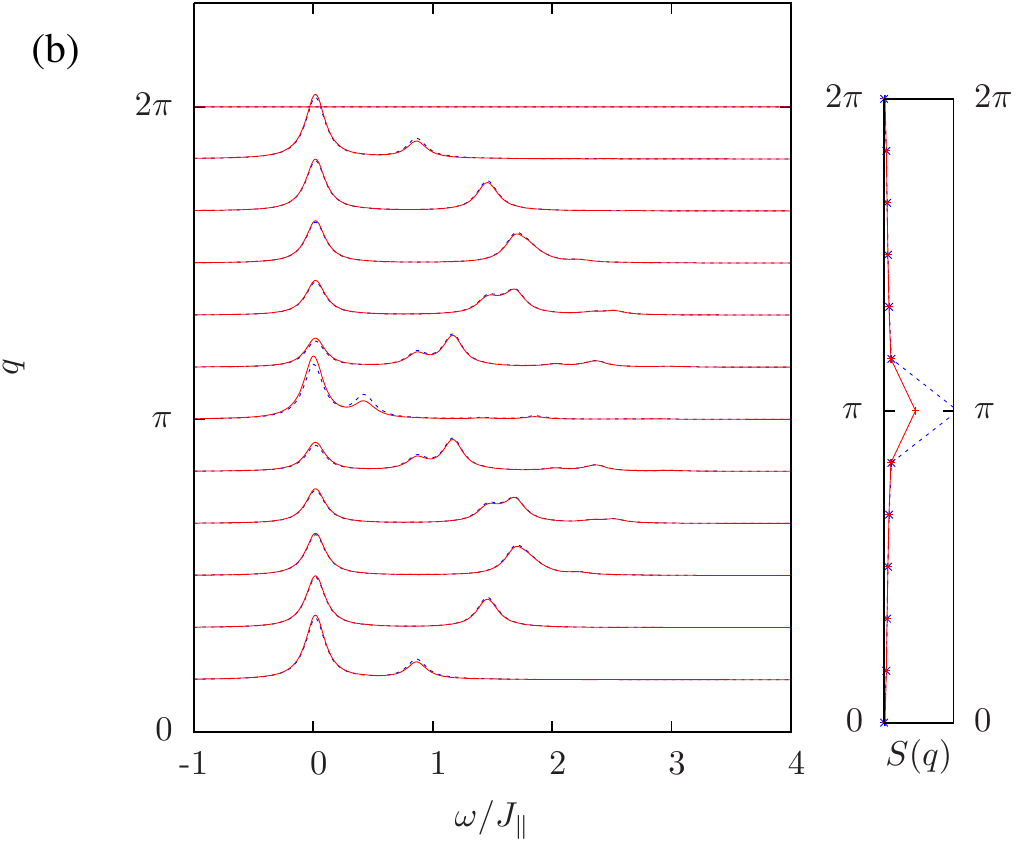}
\caption{\label{fig:EDspectra} (Color online) Dynamical spin susceptibility in the weak
         coupling limit at $J_{\bot}/J_{\|}= 0.1$:
         (a) ladder system ($\theta = 0$);
         (b) single-pole ladder system $(\theta = \pi$). In both cases, results are
            obtained on $2\times 12$ lattices with ED
            techniques. We choose a broadening $s=0.1 J_{\|}$.
            The red (solid) lines represent the bonding spectrum, $S(q,\omega)$, 
            the blue (dashed) lines corresponds to the anti-bonding spectrum, $R(q,\omega)$.
           The spectral functions in the left panels are normalized 
           to the structure factors $S(q,t=0), R(q,t=0)$, respectively; the  
           latter are shown in the right panels.  }
\end{figure}

Fig.~\ref{fig:EDspectra} presents the spin excitation spectrum for the
isotropic ladder $(\theta=0)$ and the single-pole ladder model ($\theta=\pi$) in
the weak coupling region. Precisely, it shows the dynamical spin structure
factor (depending on the momentum $q$ along the ladder), separately for
the bonding ($S(q,\omega)$) and anti-bonding ($R(q,\omega)$) configuration.
For the ladder, the dynamical spin structure factor for both bonding
and anti-bonding cases displays features of the two spinons continuum of
a single spin-$1/2$ chain~\cite{Cloizeaux1962}. 

Below we show, that such a continuum is qualitatively well
reproduced by a mean-field theory, and corresponds to the particle-hole
continuum stemming from the effective fermionic Hamiltonian.  The continuum for
bonding combination, $S(q,\omega)$, is characterized by a slightly lower energy due to the weak
ferromagnetic coupling between the chains.

\begin{figure}
 \includegraphics[width=0.4\textwidth]{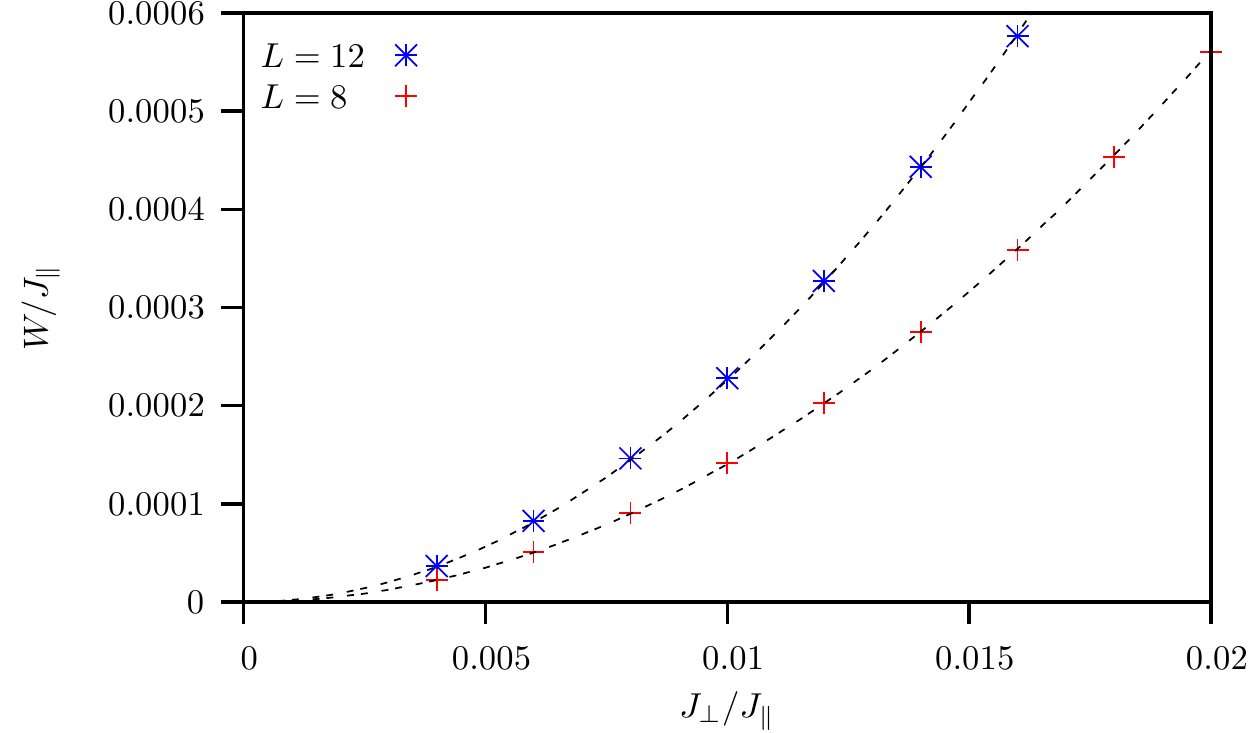}
\caption{\label{fig:EDbandwidth} (Color online)
ED results for the width $W$ in the single-pole ladder model ($\theta=\pi$). 
The effective SN interaction yields a bandwidth
proportional to $J^{2}_{\perp}/J_{\|}$.  }
\end{figure}

As apparent from Fig~\ref{fig:EDspectra}b, a narrow band emerges for the single-pole ladder
model. We define its width, $W$, as the differences in energy between the low energy 
maxima at $q=\pi$ and $q=\pi/2$  in the spin
excitation spectrum  for $\theta=\pi$. 
Associating this band with the SN splitting of dangling spins,  
 we expect the width, $W$, to  scale as
$J^{2}_{\perp}/J_{\|}$  in the weak coupling region.  This is indeed confirmed for small system size in Fig.~\ref{fig:EDbandwidth}  where the ED data is found to fit well to a $W \propto J^{2}_{\perp}$ form.

%


%

\begin{figure}[ht]
\includegraphics*[width=0.4\textwidth]{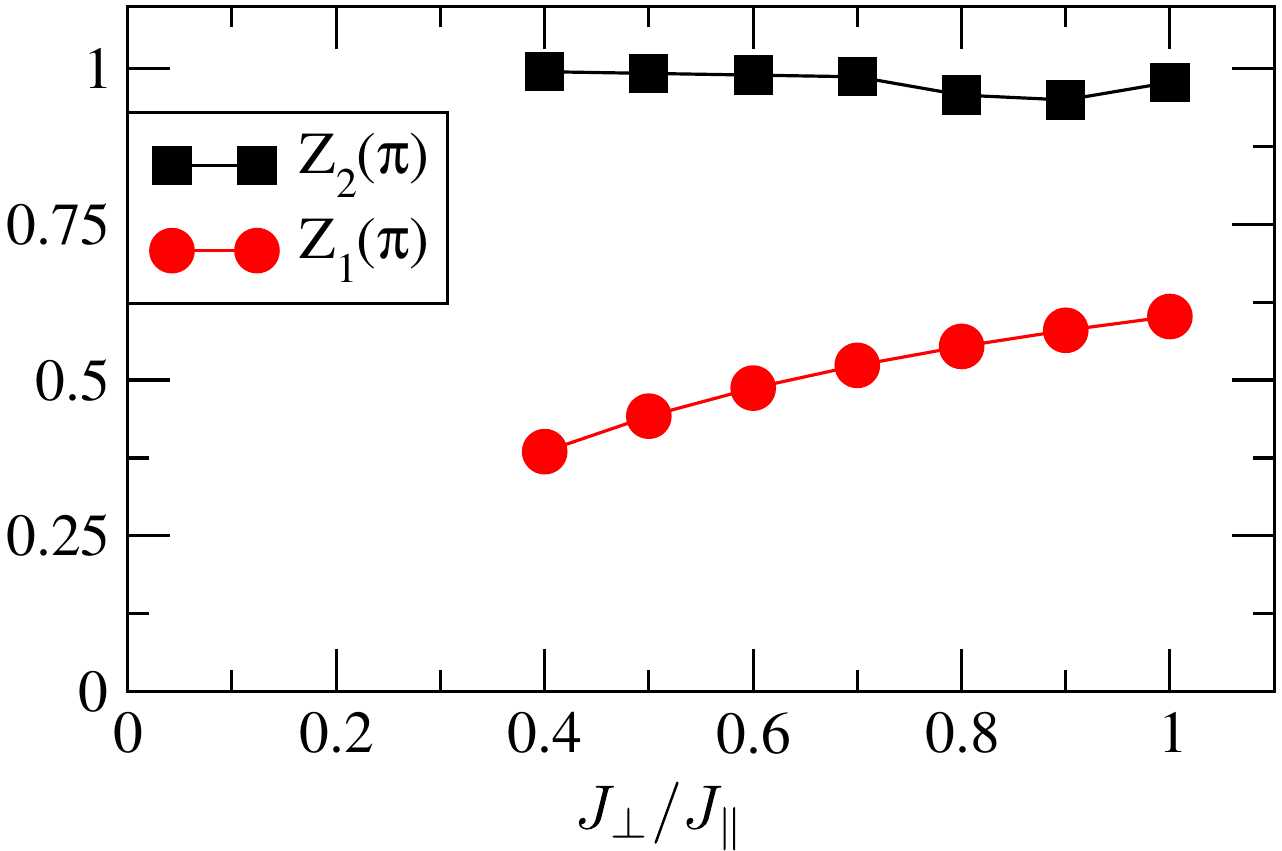}
\caption{\label{Fig:Z} (Color online) Normalized residue as a function of $J_\perp/J_\|$ for the 
ferromagnetic single-pole ladder. The calculations were carried out with ED
on an $2\times 8$ lattice.}
\end{figure}

For reasons that will be clarified below, we also investigated the  zero 
temperature spectral functions $S_{11}(q,\omega)$ and $S_{22}(q,\omega)$, Eq.\   (\ref{eq:dynspinfactor}), 
with the emphasis on the weight, or residue, of the lowest excitation energy. 
This weight is a measure of the overlap between $ \hat{S}^z_l(q) | 0 \rangle $  and the first low lying excitation 
as modeled with the effective Hamiltonian of  Eq.~(\ref{eq:Jeff}). 
 Fig.~\ref{Fig:Z}  plots  this quantity, normalized by  
$\int_0^{\infty} {\rm d} \omega S_{ll}(q,\omega)$, that is:
\begin{equation}
	Z_l (q) = \frac{ | \langle 1 | \hat{S}^z_l(q) | 0 \rangle |^2 } 
	               {   \langle  S^z_l(-q) S^z_l(q) \rangle     },
\end{equation} 
where the state $|1\rangle$ corresponds to the first magnetic excitation at wave number $q$. 
As apparent from the ED results for the single-pole ladder model (see Fig.~\ref{Fig:Z}), $Z_2(q=\pi)$ corresponding to the second leg is almost 
independent of $J_{\perp}$. Similar results are found for other momenta.  It means that nearly all the spectral weight for spins in the second leg belongs to the well -defined lowest-lying excitation. This should be contrasted with the situation of spins the first leg, where only a fraction of the spectral weight belongs 
to this model, the rest participating the spin-wave continuum. 
Clearly, the continuum fraction in $S_{11}(q,\omega)$ should increase as $J_{\perp}$ is decreased up to the decoupled situation  $J_{\perp}=0$, where one  expects $Z_2(q=\pi)=0$.

We have also computed the spin gap $\Delta$ as a function of the interleg
coupling
$J_{\perp}$ for  different values of $\theta$. 
Unfortunately,  the finite size scaling becomes difficult
in the weak coupling limit for all $\theta$ (data not shown) and an extrapolation  to the
thermodynamic limit is not feasible.   We can only confirm the differences in the scaling behavior of the spin gap
between the  the single-pole model and the ladder model, where  it is widely accepted that the gap opens linearly with
the coupling between chains ~\cite{Shelton1996,Larochelle04}. 

Concluding this subsection, we notice that the advantage of the ED method is its high accuracy, 
but the method is limited to relatively small system sizes. For the single pole ladder, 
when the gap becomes small and comparable 
to interlevel spacing $\sim J_{\|}/L$, the accuracy of the calculation becomes irrelevant.

%
%
%
\subsection{Quantum Monte Carlo, spin correlations}
\label{sec:qmc}

To extend our analysis to larger system sizes, we also used quantum
Monte Carlo (QMC) methods, performing simulations at finite inverse temperature
$\beta=1/T$. We applied two variants of the loop algorithm. For the
spin-spin correlation function and for the string order
parameter, discussed below, we used a discrete time algorithm. \cite{Evertz03}
From the spin-spin correlation function we can then extract the spectral function via stochastic
analytical continuation schemes. \cite{Sandvik98,Beach04} 
In the next subsection, we also use a continuous time loop
algorithm to directly compute the spin gap.

We start our QMC analysis with a discussion of the dynamical spin-spin correlations, which can be computed on much
larger sizes than with ED.  However, the energy resolution is limited  and hence
we can only use this approach at larger couplings than the ones reached with ED.

\begin{figure}\centering
 \includegraphics[width=0.4\textwidth]{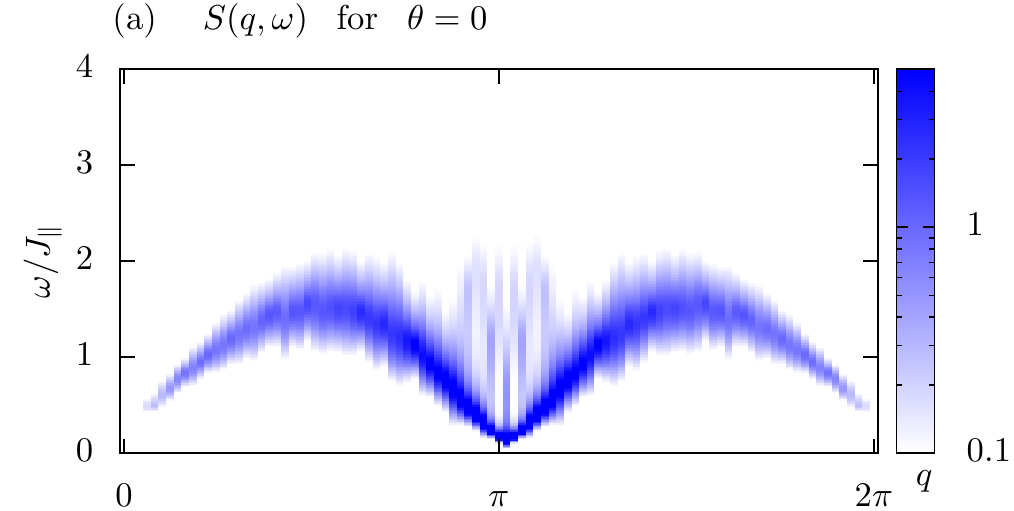}\\
 \includegraphics[width=0.4\textwidth]{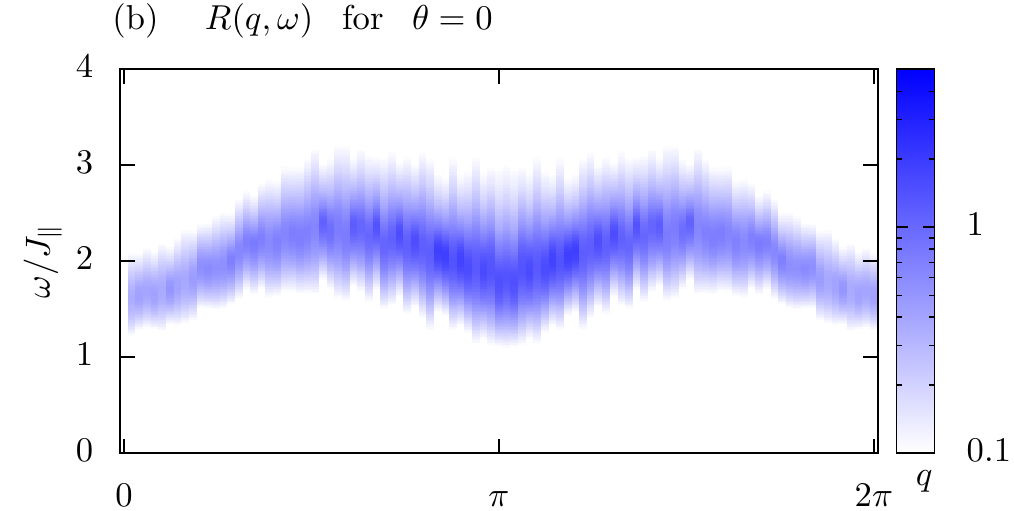}\\
 \includegraphics[width=0.4\textwidth]{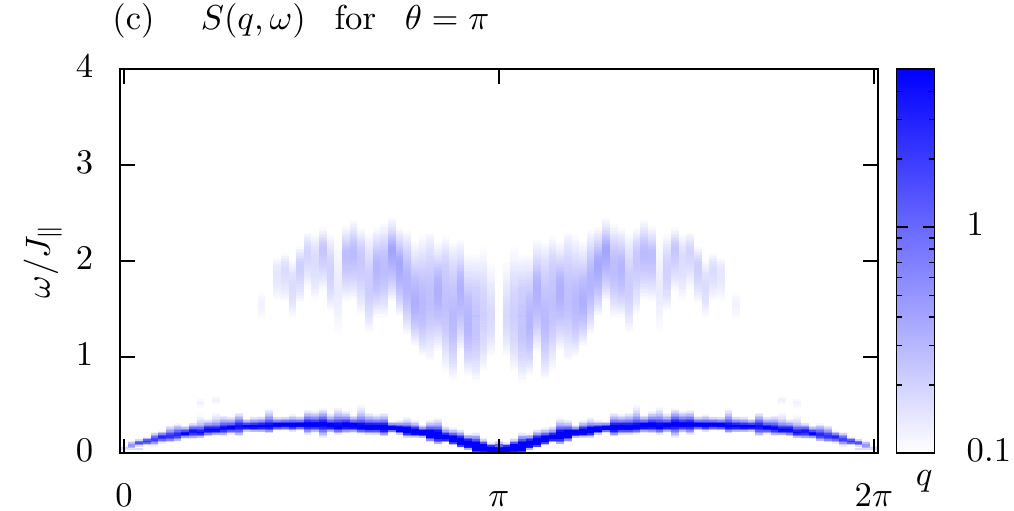}\\
 \includegraphics[width=0.4\textwidth]{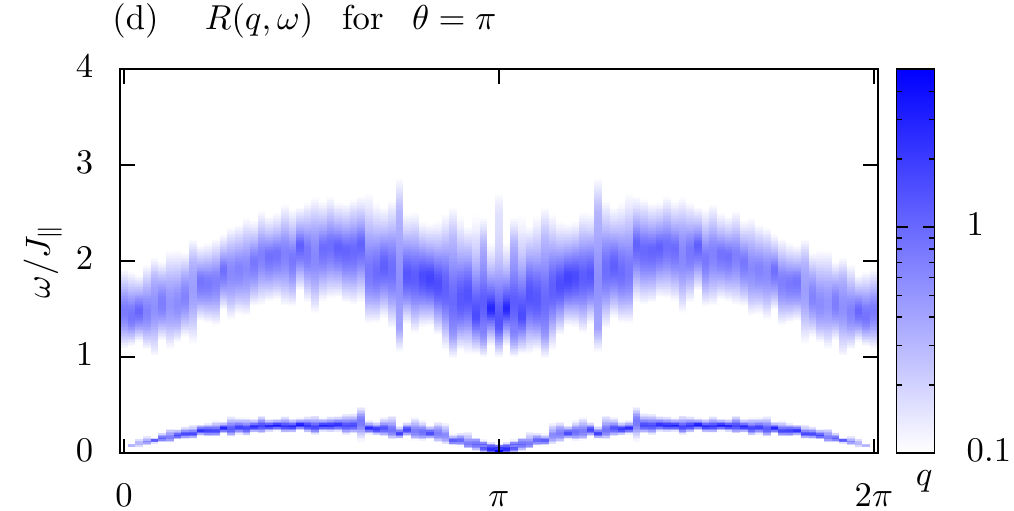}
\caption{\label{fig:QMCspectra} (Color online)
QMC results of the bonding and antibonding dynamical spin susceptibility
for the ladder ($\theta = 0$) (two top panels) and single-pole ladder ($\theta =
\pi$) (two bottom panels) systems at $J_{\bot}/J_{\|}=1.0$.
$\beta J_{\|}=200$ was taken in the simulations.}
\end{figure}

The QMC results of the dynamical spin susceptibility for $\theta=0$ (ladder) and $\theta=\pi$
(single-pole ladder) with $2\times 100$ sites at
$J_{\bot}/J_{\|}= 1.0$ ($\beta J_{\|}=200$) are shown in Fig.\ \ref{fig:QMCspectra} 
(these results were partly shown in Fig.\ 4 of Ref.\ \cite{Brunger2008}).
At $\theta =0$, inversion symmetry
$ {\mathbf{S}}_{1,i} \leftrightarrow  {\mathbf{S}}_{2,i}$
is present, such that the bonding and antibonding combinations do not mix.
Since $ {\mathbf{S}}_{i}$ is even under inversion symmetry (with respect to
the transverse direction), $S(q,\omega)$ picks up the dynamics of the triplet
excitations across the rungs.  For ferromagnetic rung couplings $J_\bot > 0$,
the low energy spin dynamics of the model is apparent in 
$S(q,\omega)$ which in the strong coupling limit maps onto the spin
structure factor of the Haldane chain. In contrast, $ {\mathbf{R}}_{i}$ is
odd under inversion symmetry and picks up the singlet excitations
across the rungs. As apparent from  Fig.\ \ref{fig:QMCspectra}b those
excitations are located at a higher  energy scale set by $J_{\bot}$ in the strong
coupling limit.  For the single-pole ladder model, $\theta =\pi$, a mixing of the bonding and
anti-bonding sectors occurs. As apparent in Fig.\ \ref{fig:QMCspectra}c,d
both $R(q,\omega)$ and $S(q,\omega)$ show high and low energy features.  
The low energy,  narrow,  dispersion curve in Fig.\ \ref{fig:QMCspectra}c  is a consequence 
of the SN interaction and reflects the slow dynamics of triplets formed across the rungs.    

\begin{figure}
\includegraphics[width=0.4\textwidth]{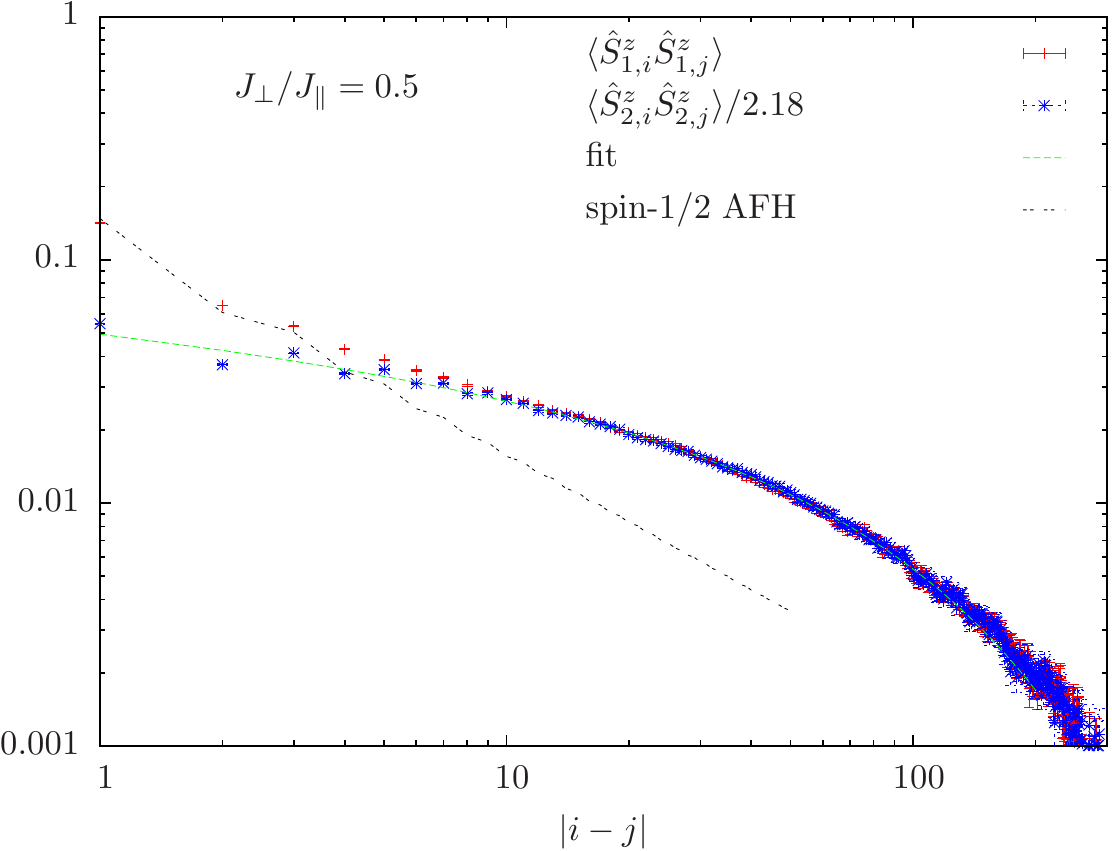}
\caption{\label{fig:QMCcorrelations} (Color online) Equal time  spin-spin correlation function
at $J_{\perp}/J_{\|} = 0.5$ for the single-pole ladder along both chains. Simulations were done 
at $\beta J_{\|}=5000$ on a $2\times800$ lattice.}
\end{figure}

In spite of the limited energy resolution of the QMC method, we can extract the value of the spin gap, by studying
the spin correlation functions at long distances. Such analysis also provides an insight of the intermediate energy 
scales. We show the behavior of the correlations  $ \langle  {S}^{z}_{1,i}  {S}^{z}_{1,j}  \rangle$ and $ \langle  
{S}^{z}_{2,i}  {S}^{z}_{2,j}  \rangle$ as function of $|i-j|$ in Fig.\  \ref{fig:QMCcorrelations} for a particular value
$J_{\perp} = 0.5 J_{\|}$ ; the distance is measured in lattice spacings. We notice that beyond 
a certain length scale both correlation functions behave similarly, differing only in overall factor. We hence plot 
in Fig.\  \ref{fig:QMCcorrelations} the function  $ \langle  {S}^{z}_{1,i}  {S}^{z}_{1,j}  \rangle$ as is, while 
$ \langle  {S}^{z}_{2,i}  {S}^{z}_{2,j}  \rangle$ is multiplied by a factor discussed below; 
after this ``normalization'' both curves are indistinguishable at large distances $d \agt 10$
(these results were partly shown in Fig.\ 3 of Ref.\ \cite{Brunger2008}).
This indicates that the lowest-energy dynamics of dangling spins $ {S}^{z}_{2,i} $ and of the main leg ${S}^{z}_{1,i}$ is 
the same slow dynamics of rung triplets, wherein these spins enter with different weight.  

At largest distances an exponential decrease of correlations is observed, corresponding to a gap in the spectral weight of 
 $ S_{jj}(q, \omega)$. In the previous section we developed a theory which accounts in a semi-quantitative way  
 the whole body of numerical findings. According to this theory, the  long distance  behavior of the 
spin-spin correlations is given by 
\begin{equation}   
	|\langle  {S}^{z}_{j,r}  {S}^{z}_{j,0}  \rangle|   =
    \frac{w^{2}_{j}}{ \pi } K_0  \left( \frac{\Delta r}{v} \right).
    \label{Eq:fit}
\end{equation}   
with   $K_0$  the modified Bessel function, $v \sim J_{\|}$ velocity of spin excitations, $w_{j}$ (with $j=1,2$) is the weight 
of the spin ${S}^{z}_{j,r}$ in the effective spin-1 variable of the single-pole ladder.  As explained above,
 we  expect $w_{j} \sim J_{\perp}/J_{\|}$ while $\Delta \sim J_{\perp} \exp (- J_{\|}/J_{\perp})$.

In Fig.~\ref{fig:QMCcorrelations} we fit the 
long-ranged equal time spin-spin  correlation function at 
$J_\perp/J_\| =  0.5 $  to this form. Several comments are in order.   

\noindent
a) Normalizing the spin-spin correlations  of the dangling spin by a factor $2.18 = w^{2}_{2}/w^{2}_{1}$  provides a
perfect agreement between the long range correlations on both legs.  Note that the numerical 
factor $2.18$ is close to $ Z_2(q=\pi)/Z_1(q=\pi) \simeq 2.25 $ as obtained 
from the data of Fig.~\ref{Fig:Z}. Hence the low energy 
dynamics of the spins on both legs are locked in together. This  observation confirms the 
picture that the low lying spin mode observed in Fig.~\ref{fig:QMCspectra}c indeed corresponds 
to the dynamics of triplets across the rungs.    \\
b) One can read off a length scale at which the functional form  of both correlation functions 
differs.  This length scale marks the crossover from high to low energy  beyond which 
an effective low energy theory can be applied.  \\
c) In our previous paper \cite{Brunger2008} we also pointed out the existence of the intermediate asymptotic,
$|i-j|^{-1/3}$, in case $J_{\perp} = 0.3 J_{\|}$. The existence of such 
power-law decay is interesting on its own, but the detailed description of this regime is beyond the scope of this 
study. This is the case when the gap becomes comparable to 
the energy spacing due to the finite size of the system, and it causes difficulties in other 
complementary numerical techniques, see e.g.\ DMRG results below.  
\\
d) For comparison we have plotted the spin-spin correlations of the spin-1/2 chain \cite{Affleck98,Essler2007} 
which, on a length scale set by the correlation length, decay much faster than the correlations of the 
single-pole ladder.  This very slow decay should be seen as a direct consequence of the {\it long  ranged} nature of the SN  interaction.  \\
e) The fit to the form of Eq.~(\ref{Eq:fit}) is next to perfect thereby providing an excellent 
description of the low energy physics. The ratio of the gap to the velocity as well as the weight of the spins in effective spin-1 variable are plotted 
in Fig.~\ref{Fig:Scales}.  In particular, assuming a linear dependence of the weight (see below), 
we  obtain  $w^{2}_{2} = 0.13 J_\perp/J_\|$ and $w^{2}_{1} = 0.060 J_\perp/J_\|$. 
The gap values are fitted by an exponential law, $\Delta \sim J_{\perp} \exp(-J_{\|}/J_{\perp})$, as discussed  in Sec.\  \ref{sec:blocks}.  
In the next subsection, we will see that this  exponential  form provides an excellent fit to the 
spin gap directly measured in QMC. The parameters of the fit shown by a dashed line in Fig.\ \ref{Fig:Scales}  
are explained in the last paragraph of the next subsection.

\begin{figure}[th] 
 \includegraphics*[width=0.4\textwidth]{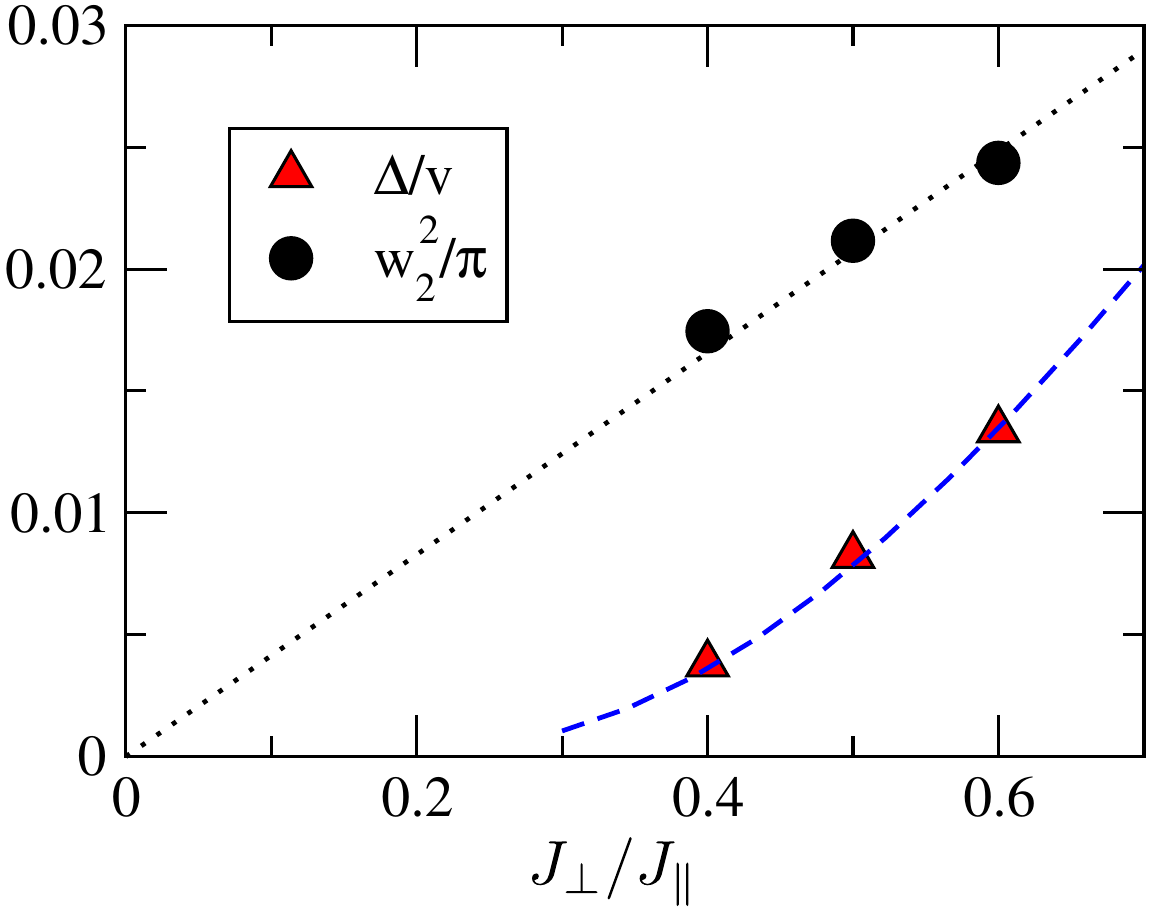}
\caption{\label{Fig:Scales} (Color online) Scales in the correlation function (\protect\ref{Eq:fit}) on the second leg 
as a function of  $J_{\perp}/J_{\|}$. The dotted line is a fit to a linear law and the dashed curve is a fit 
with $v = 0.28 J_{\|}$. The form of $\Delta$ is discussed at the end of Sec.\  \ref{sec:QMCgap}. }
\end{figure}

%
%
\subsection{Quantum Monte Carlo, spin gap}
\label{sec:QMCgap}

For the spin gap calculation the continuous-time loop algorithm of ALPS \cite{Alet05} was used.
Here, we can calculate the correlation length in imaginary time
$\xi_\tau({q})$ for a given wave vector ${q}$ via a second
moment estimator~\cite{Cooper82,Todo01}: 
\begin{equation}
\xi_{\tau}( {q}) = \frac{\beta}{2\pi}
\left(\frac{\chi( {q},\omega=0)}{\chi( {q},\omega=2\pi/\beta)}-1
\right)^{1/2}
\end{equation} 
with $\chi( {q},\omega)=\int^{\beta}_{0}d\tau e^{i \omega\tau}
\chi( {q},\tau)$ the Fourier-transform of the imaginary time dynamical
structure factor (\ref{corr-imagtime}).

 The inverse of $\xi_{\tau}( {q})$ converges
in the limit $L,\beta\rightarrow \infty$ to
$\Delta ( {q})=E_{1}({ {q}})-E_{0}$, if $\Delta( {q})\neq 0$.
Here $E_1({ {q}})$ is the minimum of the dispersion at wave vector
$ {q}$ and $E_0$ the ground-state energy: the spin gap is simply obtained
as $\Delta=\min_{ {q}} \Delta( {q})$. If the system is gapless
at $ {q}$, $\xi_{\tau}( {q})^{-1}$ is an upper bound of the
finite-size gap at $ {q}$ for any finite $L$ and $\beta$. The full
dispersion curve can therefore be calculated in principle with this method.
In practice however, the simulations suffer from large statistical errors in
$\chi( {q},\omega)$ when $ {q}$ is different
from the wave vector of the lowest lying excitation. The situation can be
ameliorated by using improved estimators~\cite{Baker95} for the imaginary time
dynamical structure factor, which is simply related to the loop sizes in the
algorithm. The wave vector picked by the loops in the algorithm corresponds to
the one given by the sign of the coupling constants~\cite{Evertz03}, which
is in our case ${  q}=(\pi,0)$ (for ferromagnetic $J_\bot > 0$ and
antiferromagnetic $J_\|< 0$). The improved estimators have a smaller variance
than conventional ones, and we therefore obtain good statistics for
$\xi_{\tau}(\pi,0)$ and the spin gap $\Delta=\Delta (\pi,0)$ when
it is finite.

\begin{figure}
\includegraphics[width=0.48\textwidth]{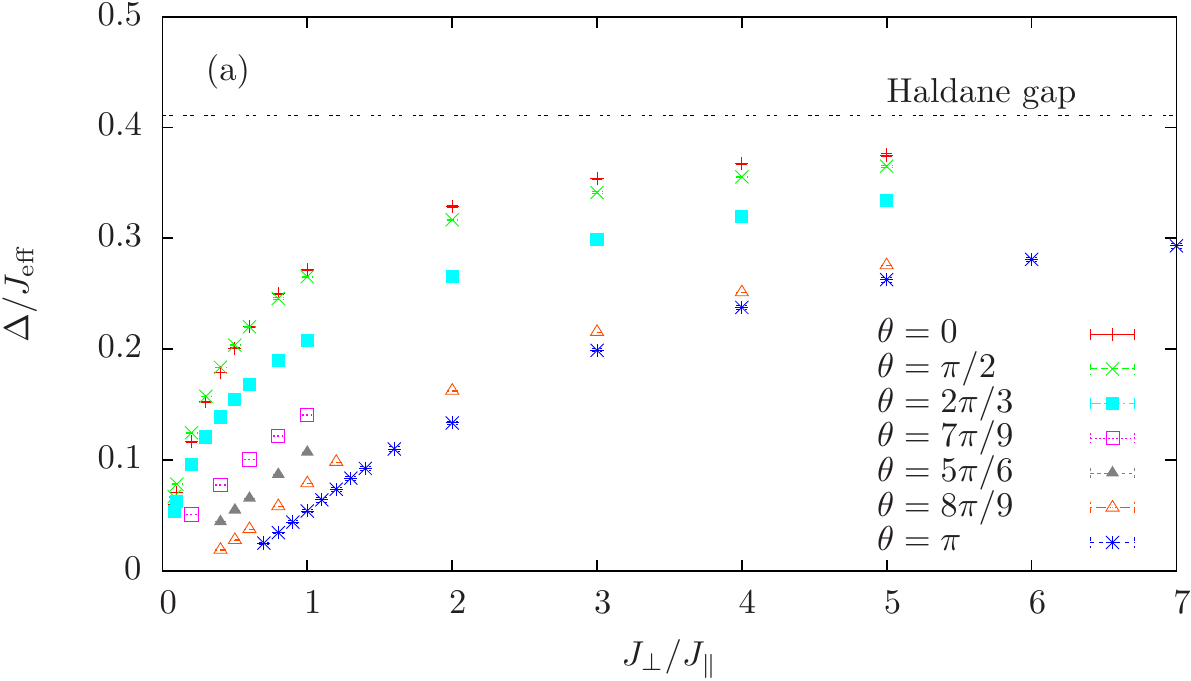}
\includegraphics*[width=0.48\textwidth]{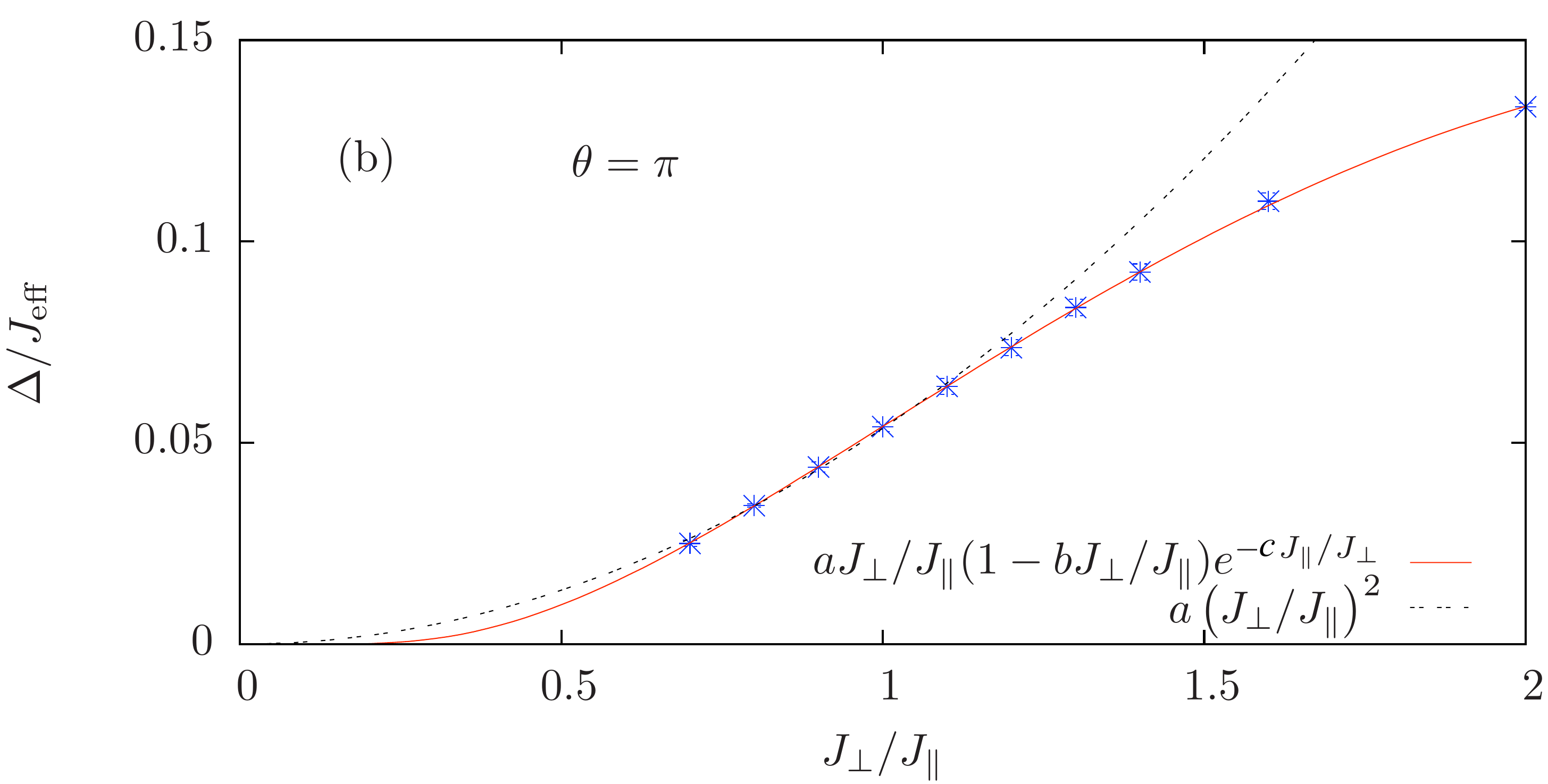}
\includegraphics*[width=0.48\textwidth]{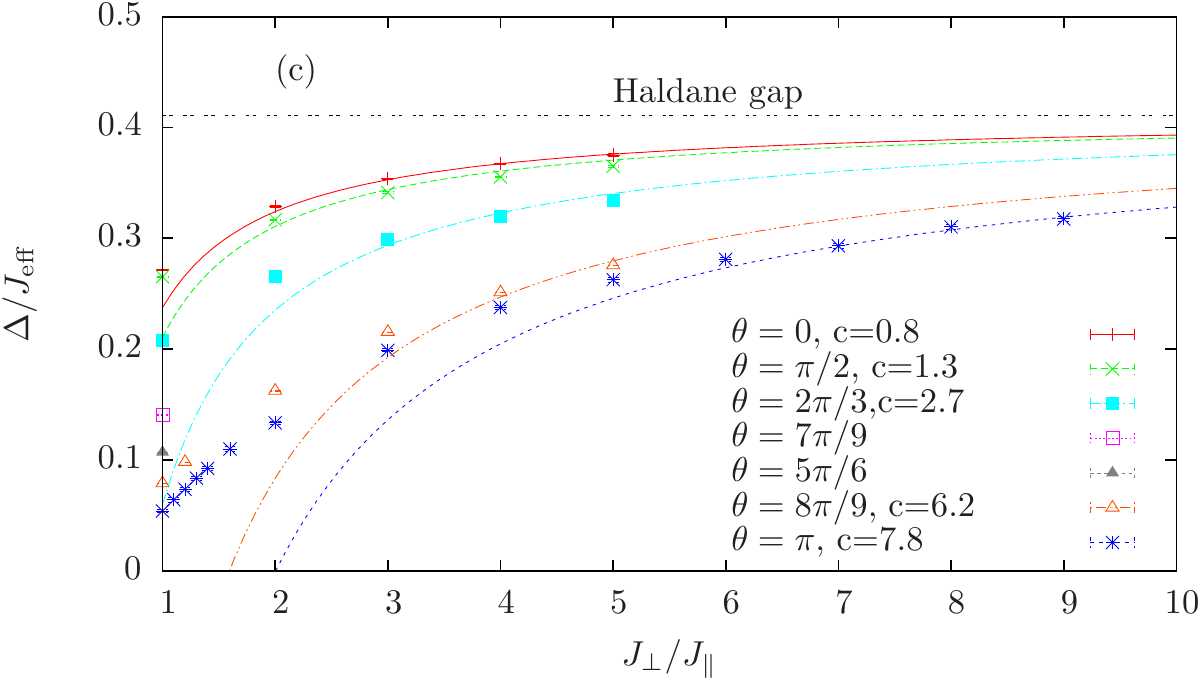}
\caption{\label{fig:QMCgap} (Color online) (a)  Size and temperature converged values of the
spin gap $\Delta$ as a function of  the coupling $ J_{\perp}/J_{\|} $ for various 
twist angles.  The gap is rescaled by
$J_{\mathrm{eff}}=\tfrac{J_{\|}}{4}\left(1+\cos^{2}(\theta/2)\right)$ such that
in the large-$J_{\perp}$-limit it converges asymptotically toward the Haldane
gap of a AF spin-$1$ chain
$\Delta_{H}/J_{\mathrm{eff}}\simeq 0.41$.
(b)  Spin gap for the single-pole ladder model, $\theta = \pi$. 
Lines denote quadratic and exponential fits for the spin gap
(see text).
(c) Spin-gap for SSHL at larger 
 $J_{\perp}$ for different $\theta$s, see detailed discussion in Sec.\   \ref{sec:app2}. 
  }
\end{figure}

Now we turn our attention to the spin gap as a
function of the coupling $J_{\perp}$ and twist angle $\theta$ (see Fig.~\ref{fig:QMCgap}a)
(these results were partly shown in Fig.\ 2 of Ref.\ \cite{Brunger2008}).

In the strong coupling limit $J_{\perp}/J_{\|}\to\infty$ the model maps onto
the AF spin-$1$ Heisenberg chain, Eq.\ (\ref{eq:Jeff}) and we expect 
  the Haldane gap $\Delta_{H}\simeq 0.41 J_{\mathrm{eff}}$.  
This behavior is clearly apparent in
Fig.~\ref{fig:QMCgap}a. However, as $\theta$ grows from $\theta=0$ to
$\theta=\pi$ the approach to the Haldane limit becomes slower. Whereas the
scaling for $\theta=0$ and $\theta=\pi/2$ are nearly similar, the scaling
behavior for the single-pole ladder ($\theta=\pi$) differs, as can be seen in Fig.\  \ref{fig:QMCgap}c.  
In Section \ref{sec:app2} we argued that at largest $J_{\perp}$ the gap should follow the law 
 $\Delta \simeq 0.41 J_{\mathrm{eff}} (1- c(\theta) J_{\mathrm{eff}}/{|J_\perp|})$. 
 Now we confirm this picture by fitting the numerical data with this form at the largest $J_{\perp}$, as shown in 
Fig.\ \ref{fig:QMCgap}c. We see that $c(\theta)$ definitely grows as a function of $\theta$ and reaches 
a large value $c(\theta) \simeq 8 $ for the extreme case of the single-pole ladder, $\theta=\pi$. 
 In the intermediate region,  $1\alt J_{\perp}/J_{\|} \alt c(\theta)$,  we cannot 
expect a good fit of the gap, as clearly visible in Fig.\ \ref{fig:QMCgap}c.   

 For the ladder system ($\theta=0$) our data stand in agreement with the independent QMC
calculations of Ref.~\onlinecite{Larochelle04};  the spin gap opens linearly
with respect to the coupling $J_{\perp}$ up to logarithmic corrections.
It is beyond the scope of this work to pin down the exact form of
the logarithmic corrections, and we refer the reader to
Ref.~\onlinecite{Larochelle04} for further discussions.  This behavior is stable up to 
to large twist angles, and it is only in the very close vicinity of $\theta = \pi$ that a different behavior is observed.  

As discussed in the previous section,  
the spin gap at $\theta = \pi$ is expected to decrease exponentially with decreasing values of $J_\perp$, as 
$\propto J_{\perp} \exp(-J_{\|}/J_{\perp})$. In order to fit the QMC data of Fig.\ \ref{fig:QMCgap}b in a  wider region, $J_{\perp} \alt J_{\|}$,
we also allow for a correction in the prefactor in this law, namely 
we assume the dependence $\Delta = a J_{\perp} (1- b J_{\perp}/J_{\|}) \exp(-c J_{\|}/J_{\perp}) $. 
Fitting the data  of Fig.\ \ref{fig:QMCgap}b  to this form gives $a=0.077, b= 0.32, c= 1.34$. 
These parameters  can now be used to fit the data for the gap, $\Delta/v$, as extracted from the spatial decay of correlations
 in Fig.\  \ref{Fig:Scales}. The only adjustable parameter there is $v/J_{\|}$, and we obtain a good agreement in Fig.\ \ref{Fig:Scales} for $v = 0.28 J_{\|}$. 
 The effective model of the previous section leads to an expression $v=4w^{2}_{1}\xi J_{\|} $.  
 Comparing these values, we determine the crossover scale $\xi = 1.2 J_{\|}/J_{\perp}$, which separates the
 long-distance behavior from the short-distance one. For the particular value $J_{\perp} = 0.5 J_{\|}$, we obtain 
 $\xi \sim 3$, and this scale is clearly visible in the merging of curves for the spatial dependence of correlations in 
 the first and second legs, as shown in Fig.\ \ref{fig:QMCcorrelations}. 
 We further show in Fig.\ \ref{fig:QMCgap}b the result of a quadratic fit to the gap
$\Delta \sim (J_\perp / J_\|)^2$. While such a fit is reasonable for
low $J_\perp$ as remarked in Ref.\  \onlinecite{Brunger2008}, the above exponential form appears
to provide a better fit in a larger $J_\perp$ window.

%
%
\subsection{String Order Parameter}
\label{sec:SOP}
To pin down the nature of the ground state, and in particular for the single-pole ladder, 
we  compute the string order parameter  characterizing the Haldane phase.
In the strong coupling region the system maps onto an effective spin-$1$ chain,
for all twist angles $\theta$ and the ground state can  viewed 
in terms  a valence bond solid (VBS)~\cite{Affleck1987,Affleck1988,Schollwock1996}. In the
VBS state the spins on a rung form triplets in such a way that neglecting 
triplets with z-component of spin $m=0$ reveals a  N{\'e}el order.  
This {\it hidden} AF order,  is characteristic of the  Haldane
phase and, as shown by
Nijs and Rommelse~\cite{Nijs1989},  is revealed by the
non-local string order parameter
\begin{eqnarray}
\mathcal{O}_{s}=\Big{\langle} {S}^{z}_{n_{0}}
\exp\left[i \pi\sum^{n_{0}+L/2-1}_{j=n_{0}} {S}^{z}_{j}\right]
 {S}^{z}_{n_{0}+L/2}\Big{\rangle}
\end{eqnarray} 
where $ {S}^{z}_{i}= {S}^{z}_{1,i}+ {S}^{z}_{2,i}$, $n_{0}$ stands for
an arbitrary rung and $L$ denotes the system length. $\mathcal{O}_{s}$ is also
sensitive to a {\it true} AF order. To distinguish between a hidden AF order
and a {\it true} N{\'e}el order another order parameter has to be
introduced~\cite{Nijs1989} 
\begin{eqnarray}
\mathcal{O}_{H}=\Big{\langle}
\exp\left[i\pi\sum^{n_{0}+L/2}_{j=n_{0}} {S}^{z}_{j}\right]
\Big{\rangle}
\end{eqnarray} 
which is zero in the Haldane phase (hidden AF order) and finite in the N{\'e}el
phase. Starting from the strong coupling region where the system is clearly in
the Haldane phase, $\mathcal{O}_{s} \ne 0$ and $\mathcal{O}_{H}=0$,
we analyze the evolution  of the string order parameter as a function of  the coupling
$J_{\perp}$ and the twist angle $\theta$. For $\theta=0$ the order parameter
$\mathcal{O}_{s}$ stays finite and $\mathcal{O}_{H}$ is zero for all couplings.
Hence the ladder system remains in the Haldane phase, independent of
$J_{\perp}$. 

%

\begin{figure}
\includegraphics[width=0.45\textwidth]{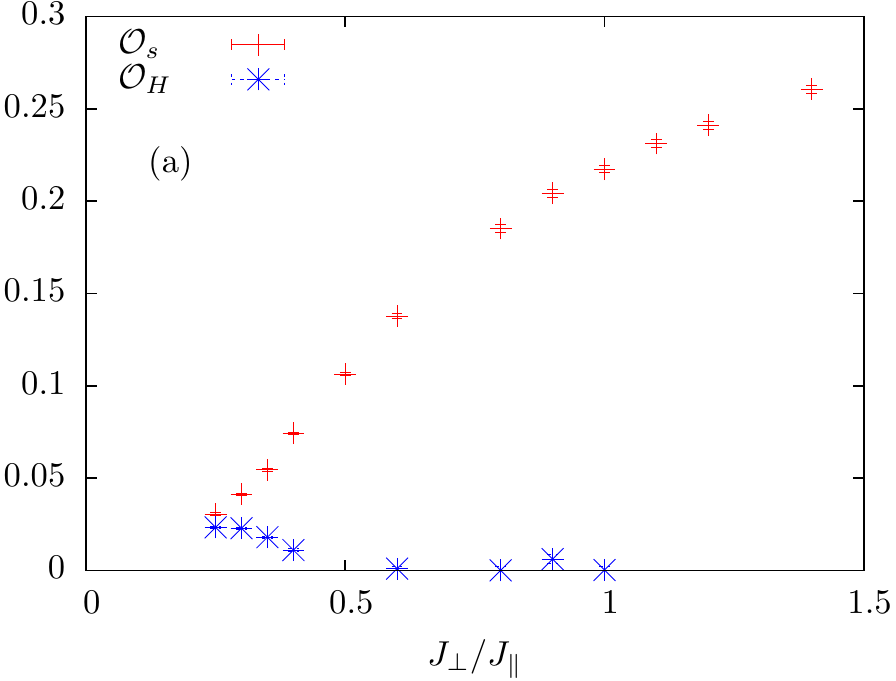} \\
\includegraphics[width=0.45\textwidth]{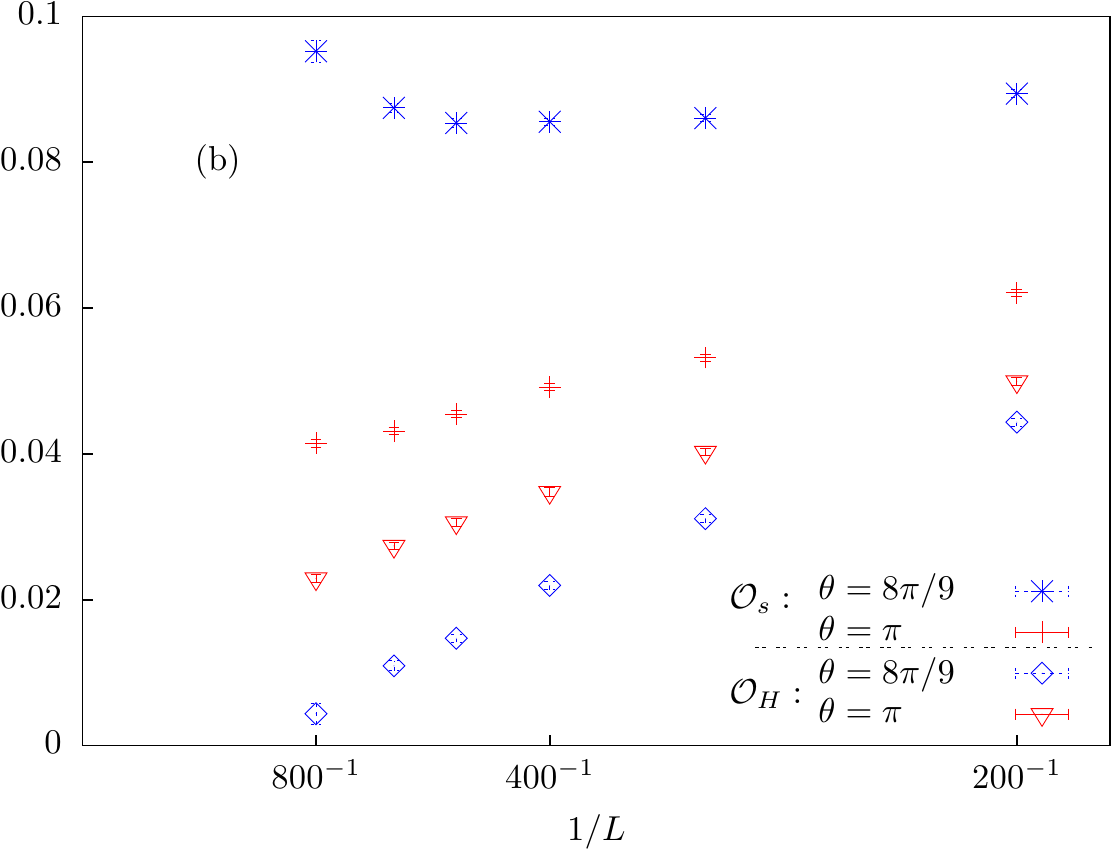}
\caption{\label{fig:QMCstringorder} (Color online) (a) String order parameters
$\mathcal{O}_{s}$ and $\mathcal{O}_{H}$ as a function of interleg coupling
$J_{\perp}$ for $\theta=\pi$. In the parameter range $J_{\perp}/J_{\|} <0.5$ finite size
effects are still present. (b) Finite size scaling of the order parameters for
$\theta=8\pi/9$ ($J_{\perp}/J_{\|}= 0.2$) and  $\theta=\pi$
($J_{\perp}/J_{\|}= 0.3$). The simulations are carried out up to
$\beta J_{\|}=7000$ and $2\times 800$ spins.}
\end{figure} 

The situation for the single-pole ladder is much more
delicate (see Fig.~\ref{fig:QMCstringorder}a, 
these results were shown in Fig.\ 5 of Ref.\ \cite{Brunger2008}).
For $J_{\perp}/J_{\|} <0.4$ the string order parameter $\mathcal{O}_{H}$
appears to be non-zero, thus indicating N{\'e}el order. In the previous
section we have shown, that at weak couplings and for $\theta=\pi$ the SN
interaction generates a very slow decay of the spin correlations. For instance, the
correlation length $\xi$ for $J_{\perp}/J_{\|}= 0.3$ is larger than the system
size and the very slow decay of the spin-spin correlation functions  at
distances smaller than $\xi$ mimics AF order.  However, when increasing system size, one expects $ \mathcal{O}_{H}$  to vanish  and  $
\mathcal{O}_{s}$ to converge to a finite value. This expectation is supported 
by  a  finite size scaling of the string order parameters, as shown in Fig.~\ref{fig:QMCstringorder}b
 for $\theta=8\pi/9$ and $\theta=\pi$.   The crossover between the AF order for
small systems and the disordered phase is rather obvious
 for $\theta=8\pi/9$   ($\cos ^{2}\tfrac{\theta}{2} \simeq 0.03 $)  at 
$J_{\perp}/J_{\|}= 0.2$. For small lattice sizes the system seems
to be in the AF ordered phase as indicated by the fact that both string order
parameters are finite. However, for increasing lattice sizes the order
$\mathcal{O}_{s}$ remains nearly constant, whereas the order parameter
$\mathcal{O}_{H}$ decreases and  finally vanishes. At this point the
order parameter $\mathcal{O}_{s}$ increases again. 
For the single-pole ladder ($\theta=\pi$) at $J_{\perp}/J_{\|}= 0.3$ we do not observe the disappearance  
of $\mathcal{O}_{H}$, which shows that finite size effects are strong in this case even for a system as large as $L=800$.

\subsection{DMRG analysis}
\label{sec:DMRG}

\begin{widetext} 

\begin{figure}[ht]
\includegraphics[width= 0.75\textwidth 
 ]{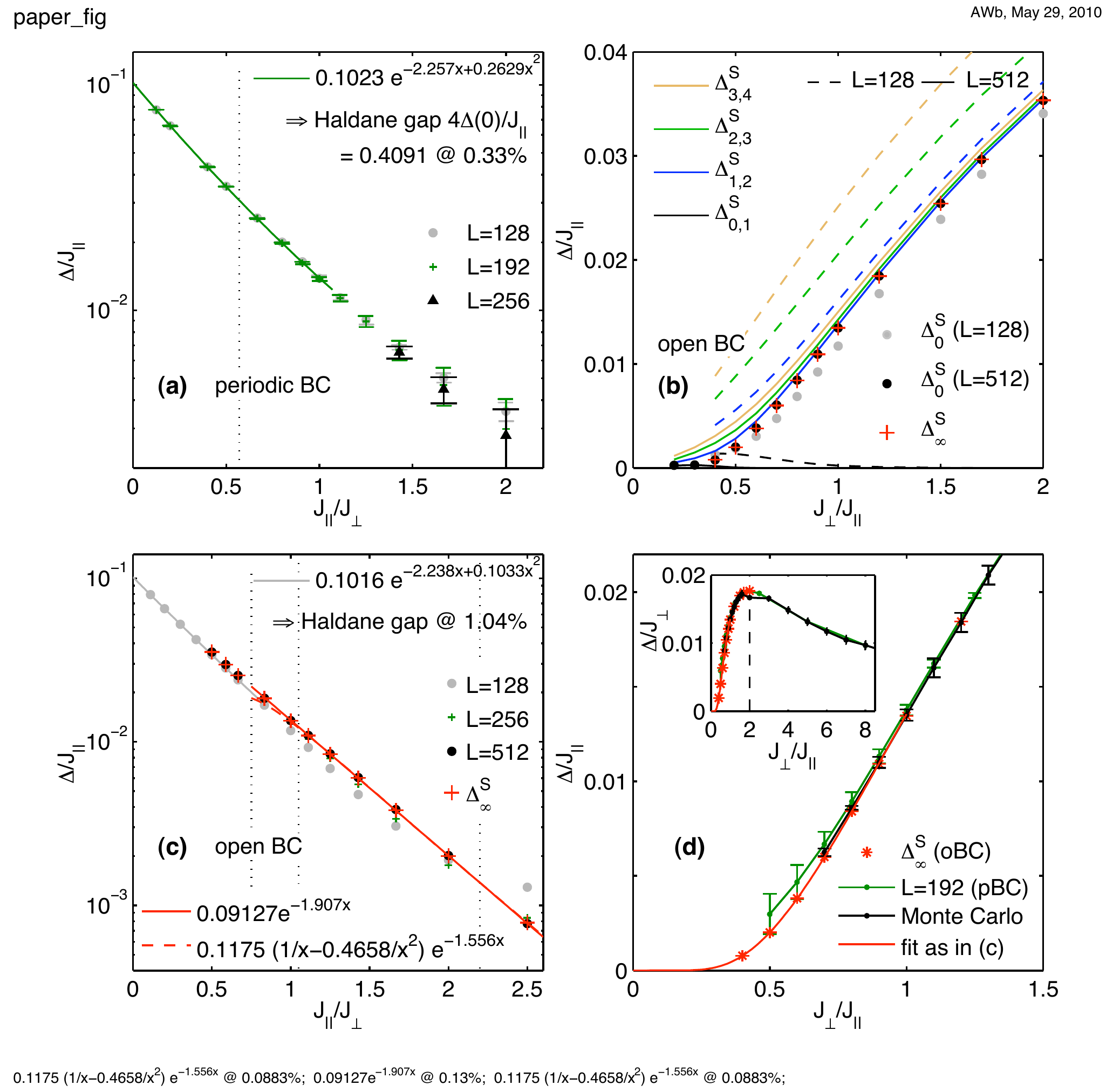}
\caption{(Color online)
DMRG analysis of the single-pole ladder ($\theta=\pi$) --
Panel (a) shows the spin gap between the ground states of the $S_z=0$
and $S_z=1$ spin sectors for several system sizes using periodic
BCs (double chain, see text). The error bars indicate
the convergence with respect to the number of states $D$ kept (with an
extrapolation in $1/D \rightarrow 0$ with $D\le
5120$).  The convergence to the Haldane gap for
$J_\perp \gg J_{\|}$ is reproduced within 1\%
relative error, within the fitting-range at small $J_{\|}/J_\perp$
indicated by the vertical dotted line.
Panel (b) shows spin gap data using open BCs
for the ground states of several spin sectors $S_z \in \{0,1,2,3,4\}$.
Keeping up to 2560 states, all energies are converged with
negligible uncertainties. The consecutive energy differences (spin
gap) $\Delta^S_{S_z,S_z+1}$ between the ground states of the
spin sectors $S_z$ and $S_z+1$ are plotted vs. $J_\perp/J_\Vert$
for $L=128$ (dashed) and $L=512$ (solid), respectively.   The vertical
order (top to bottom) of the entries in the legend matches the order
of the lines appearing in the panel (top to bottom). The extrapolated
spin gap $\Delta^S_0$ (grey and black dots for $L=128$ and $L=512$,
respectively), as well as $\Delta^S_{\infty}$ (red crosses), as obtained from a
finite size scaling of $\Delta^S_{1,2}$ for systems of length $L \in
\{128,256,512\}$, are shown.
Panel (c) shows $\Delta^S_0$ from panel (b) for $L \in \{128,512\}$
and also for $L=256$ together with $\Delta^S_{\infty}$ (all
for open BC) on a semilog plot again with inverted x-axis
similar to panel (a). The extrapolated value of the  
gap for small $J_{\|}/J_\perp$ for open BC compares well
with the exact Haldane gap within 1\% relative error already 
 for the smaller system size $L=128$ (fit range
up to first vertical dotted guide). Fits to
the data for large $J_{\|}/J_\perp$ are
shown in red solid and dashed curves.
The fit range chosen is again indicated by the two vertical dotted
guides, here for $J_\Vert/J_\perp > 1$. 
Panel (d) summarizes the data obtained with periodic (green solid) and open (red asterisk) boundary conditions 
DMRG calculations, as compared with QMC data (black solid). The red solid line is a fit for large
$J_{\|}/J_\perp$ similar to the one in panel (c). Error-bars indicate the respective uncertainty in energy. 
Inset shows that $\Delta/J_{\perp}$ reaches the maximum at $J_\perp \simeq 2 J_\|$.
}%
\label{fig:dmrg}%
\end{figure}

\end{widetext}

In this section we present our analysis of 
the Hamiltonian, Eq.\ (\ref{eq:SSHL-hamilton1}), with the use of the Density Matrix 
Renormalization Group (DMRG) \cite{White92} for the single-pole
ladder, i.e.\   $\theta=\pi$. Overall, our results presented in
Fig.\ \ref{fig:dmrg} clearly support a non-analytic exponential
scaling of the gap in $J_\perp$ as suggested by the analytical
considerations in this paper.

For the calculation of the spin gap the specific choice of the
boundary condition (BC) is crucial. While DMRG prefers open BCs for
numerical stability and accuracy, they must be dealt with carefully
in order to separate boundary effects from bulk effects.
Periodic boundary conditions can be applied within DMRG,
\cite{Verstraete2004,Pippan10}  yet with somewhat limited accuracy and
efficiency. In this paper we therefore adhere to the conventional
DMRG with open BC. In order to still deal with a Hamiltonian with
periodic BC, a long bond connecting the ends of the chain can be
introduced for short system sizes. Alternatively for somewhat longer
system sizes, the chain with periodic boundary can be reshaped into a
double chain with ends connected to form a loop. We adopted the
latter approach since it is stable in finding the ground state of
the system. Nevertheless it is enormously
costly numerically, and keeping up to 5120 states for chain lengths
up to $L=256$ rungs total, the DMRG results still showed significant
uncertainties in the ground state energy for small $J_\perp/J_\|$
insufficient to accurately resolve the exponentially small gap 
(see Fig.~\ref{fig:dmrg}, panel a).

When compared to the other calculations (see, e.g., panel (d) below),
the gap for periodic BC  is consistently
overestimated for small $J_\perp/ J_\Vert$. 
This comes from the fact that the ground state is
typically well-represented (smaller block entropy due to gap), while
the excited state at larger $\SZ$ has a larger block entropy as it is a
part of the continuum. Due to the limited number of states kept,   
the excited state is less accurately represented which leads to an overestimated gap. This effect is more
pronounced for large system sizes as can be clearly also seen from
Fig.~\ref{fig:dmrg}a.

At the same time the numerical data for the gap for large $J_\perp/ J_\Vert$ 
are reliable, allowing for an extrapolation towards the known value of 
the Haldane gap with less than 1\% relative error as shown in panel (a).
The fitting-range used was $J_\Vert/J_\perp \le 0.55$ as indicated by the
vertical dotted guide in the panel.  With a fit of the form
$\Delta(J_\perp) \sim e^{-\const J_\Vert / J_\perp}$ this shows that
the Haldane gap is reached rather slowly when increasing $J_\perp$.

With periodic BC being of limited accuracy as explained above,
we adopted the plain open BC also on the level of the Hamiltonian for the rest
of the DMRG calculations. Hence we deal with a single ladder, in
contrast to the connected double ladder above. Keeping up to 2560
states  leads to clearly converged numerical data for all
$J_\perp/J_\Vert$ analyzed. Note nevertheless, that the block entropy
rises rapidly for $J_\perp/J_\Vert < 0.5$ due to the near degeneracy
of the dangling spins on the single-pole ladder in the limit
$J_\perp/J_\Vert \rightarrow 0$.

 From the numerical analysis for systems with open BC,  one observes spin 1/2
edge excitations  visible in terms of an
alternating finite $\langle S_{z,i} \rangle$ along the sites $i$
which decay exponentially with the distance from the ends of
the chain. \cite{Kennedy1990}
As the total spin $\SZ$ is a conserved quantum number, the individual
symmetry sectors can be analyzed separately, with the overall ground
state lying within $\SZ=0$. However, as it turns out, the
$\SZ=0$ sector has clear $S_z=\pm\frac{1}{2}$ edge
excitations with opposite signs  being
consistent with $\SZ=0$. The situation is similar in the $\SZ=1$ sector, but this time yet
with equal signs of alternation, consistently adding up to $\SZ=1$. Therefore
$\SZ=0$ and $\SZ=1$ are degenerate in the thermodynamic limit
yielding a four-fold degenerate ground-state due to the presence of
the open boundary~\cite{Kennedy1990}. 
Only starting with $\SZ=2$ a true bulk excitation
is generated. \cite{Almeida2007}. In order to extract the spin gap, we therefore prefer to monitor the splitting
in energy with respect to $\SZ \ge 2$.
To this end the energy differences $\Delta^S_{S_z,S_z+1}$ between the
ground states of the consecutive spin sectors $S_z$ and $S_z+1$ are
calculated and plotted in Fig.~\ref{fig:dmrg}b. From the above argument,
$\Delta^S_{0,1}$ clearly vanishes in the thermodynamic limit as the
overlap between the spin 1/2 excitations confined to the boundaries decays
exponentially with system size. Therefore only
$\Delta^S_{S_z,S_z+1}$ for $S_z>0$ resembles the gap with a
finite-size effect added with each increment of $S_z$.
Note in Fig.~\ref{fig:dmrg}b that all curves for
$S_z>0$ start falling onto the same line as $L$ increases. This indicates, 
consistently with the notion of a spin gap, 
that for large enough system sizes 
each increment of $\SZ$ by 1 costs an energy equal to the gap value.
Note, however, that the lowest energy states for the $S_{z}>1$ sectors may already lie within 
a continuum of states.

The strategy then to extract the spin gap is two-fold: (i) extrapolation of $\Delta^S_{S_z,S_z+1}$ for 
$S_z>0$ for constant system size $L$, using a quadratic
fit towards $S_z=0$ to eliminate finite-size effects with increasing
$S_z$,
\begin{equation}
\Delta^S_0 \equiv \lim_{S_z\rightarrow 0} \Delta^S_{S_z,S_z+1}, 
\quad (L=\mathrm{const}) , 
\label{def.Delta0}
\end{equation}
(grey and black dots in Fig.~\ref{fig:dmrg}b), and (ii) actual finite-size scaling on $\Delta^S_{1,2}$, i.e. the lowest $S_z$
that yields a finite spin gap $\Delta^S_{S_z,S_z+1}$ in the
thermodynamic limit,
\begin{equation}
\Delta^S_\infty \equiv \lim_{1/L \rightarrow 0} \Delta^S_{1,2}
\label{def.DeltaInf}
\end{equation}
(red crosses Fig.~\ref{fig:dmrg}b). As can be seen in the same
panel, both strategies nicely agree with each
other for $L=512$, indicating that the data are well converged and consistent. It also shows 
that Eq.\ (\ref{def.Delta0}) is a valid way of extracting the spin
gap for systems that are large enough.

In order to analyze the data at small $J_\perp /
J_\Vert$, the data for $\Delta^S_0$ and $\Delta^S_\infty$ are
(re)plotted in Fig.~\ref{fig:dmrg}c with inverted x-axis on a
semilog-y plot, as we expect a non-analytic behavior of the form
$\Delta \sim e^{-\const J_\Vert / J_\perp}$ (and as motivated by
previous studies on spin ladders too \cite{White96}). As a consistency check
for open BCs, the gap $\Delta^S_0$ for $L=128$ is again extrapolated
towards $J_\Vert/J_\perp \rightarrow 0$ to retrieve the Haldane gap
with reasonable accuracy of 1\%.

For large $J_\Vert/J_\perp > 1$, the gap $\Delta^S_\infty$ in
the thermodynamic limit (red crosses in Fig.~\ref{fig:dmrg}c) can be
fitted nicely using exponential forms of either type
\begin{subequations}
\begin{eqnarray}
\Delta(J') &=& a_0 e^{-{c_1}/{J'}} \text{, and} \label{fitfun.a} \\
\Delta(J') &=& \left(a_1 J' - a_2 (J')^2\right) e^{- {c_2}/{J'}} \label{fitfun.b}
\end{eqnarray}\label{fitfun}
\end{subequations}
with $J' \equiv {J_\perp}/{J_\Vert}$, and $a_i$ and $c_i$ being
fitting parameters (see Fig.~\ref{fig:dmrg}c where $x \equiv 1/J'$). Note that the $(J')^2$
term in Eq.~(\ref{fitfun.b}) is important to obtain a clear agreement with
the numerical data.The data are equally well fitted by both forms in the region $J_\Vert / J_\perp > 1$, with deviations of the fit
in Eq.~(\ref{fitfun.b}) (dashed line in Fig.~\ref{fig:dmrg}c) visible only
outside the fitted region, at $ J_{\|} < J_{\perp}  $. 
Note also that despite the fact that the fitting-range chosen to be
$J_\Vert / J_\perp \in\ [1.05,2.25]$ (as indicated by the
vertical dotted guides in Fig.~\ref{fig:dmrg}c), both fits
extrapolate well up the last data point at 2.5.

Finally, the DMRG results for the gap are summarized and
directly compared to the QMC simulations in
Fig.~\ref{fig:dmrg}d, as a function of $J_\perp / J_\Vert$. The gap
$\Delta^S_\infty$ in the thermodynamic limit (red asterisks) clearly
lies within the error bars of the other less accurate data sets,
while its own error bars are negligible. The results obtained this way are then reliable down to smaller $J_\perp / J_\Vert$. The
exponential fit reproduced from Fig.~\ref{fig:dmrg}c (solid red line)
in panel (d), finally, illustrates the extremely fast decay of the gap towards small $J_\perp$. Yet as clearly
supported by Fig.~\ref{fig:dmrg}c, the spin gap remains finite in this region.
The inset in Fig.~\ref{fig:dmrg}d illustrates that, when scaled to $J_{\perp}$,  
the gap in the single-pole ladder has a maximum at $ J_{\perp} \simeq 2 J_{\|}$ and decreases exponentially 
at smaller $J_{\perp}$. This is to be contrasted with the symmetrical ladder where 
 $\Delta / J_{\perp}$ saturates to a constant at $J_{\perp}\to0$.

\section{Conclusions}
\label{Sec:Conclusions}

In conclusion, we investigated asymmetric spin ladders, with different values of exchange interaction
of spins $S=1/2$ along the two legs as parametrized by $\theta$. For ferromagnetic rung coupling  
$J_{\perp}$ the spectrum of excitations is characterized by a Haldane gap, as expected for the effective spin-1
rung variables which are coupled antiferromagnetically along the chains. We confirm this by the numerical analysis of the spin 
gap, spin correlation functions and of the corresponding string order parameter. 

The most intriguing behavior is observed near the single-pole situation, {\it i.e.} in the absence of 
exchange along the second leg. Our extensive numerical analysis shows that  the 
spin gap decreases with $J_{\perp}$ exponentially fast, $\Delta \sim J_{\perp} \exp( - J_{\|}/J_{\perp})$, unlike
the conventional symmetric ladder behavior, $\Delta \sim J_{\perp}$. In order to explain the whole body of numerical 
data, we develop a theory, which takes into account the indirect Suhl-Nakamura interaction between spins
 and the formation of large effective blocks
of spins. In a certain sense, the formula for the gap as obtained from this approach combines 
the ``quantum'' prefactor $J_{\perp}$ and 
the semi-classical exponent $\exp( -J_{\|} / J_{\perp})$ arising from the large-block picture.    

In summary, we have a strong evidence 
that a spin gap opens for any $J_{\perp}$ but it may be exponentially suppressed nearly
single-pole situation. 
This smallness leads to difficulties in the actual observation of this gap, both due to the finite resolution of the tool 
employed (numerical or experimental) and to the finite-size effects associated with a large correlation length. 

The case of a negligible gap was reported in Ref.~\onlinecite{Brunger2008} for $J_{\perp}=0.3 J_{\|}$.  
We did not discuss this situation in the present paper, since at these parameters the
correlation length is larger than the system size and the true string order parameter is not formed, cf.\ 
Fig.\ \ref{fig:QMCstringorder}. In such case the observed power-law decrease of correlation functions should 
be viewed as an intermediate asymptote, and the origin of the particular value of the decay exponent 
reported in  \cite{Brunger2008} is not clear.  
 
\begin{acknowledgments}
We thank A.A.\ Nersesyan, K. Kikoin, F.H.L.\ Essler, P.\ Schmitteckert and
S.R.\ White for fruitful discussions. 
The work of D.A.\ was supported in part by the RFBR Grant.  
A.W. acknowledges financial support from DFG (De 730/3-2, SFB-TR12) and the
Excellence Cluster Nanosystems Initiative M\"unich (NIM).
C. B. and F.F.A. thank the DFG for financial support under the grant numbers: AS120/4-2 and
 AS120/4-3.  Part of the numerical calculations were carried out  at the LRZ-M\"unich, Calmip (Toulouse) as
well as at the JSC-J\"ulich. We thank those institutions for generous allocation of CPU
time.
\end{acknowledgments}
  
\appendix

\section{Jordan-Wigner Mean-Field Approach}
\label{sec:meanfield}
\begin{figure}[btp]
 \includegraphics[width=0.45\textwidth]{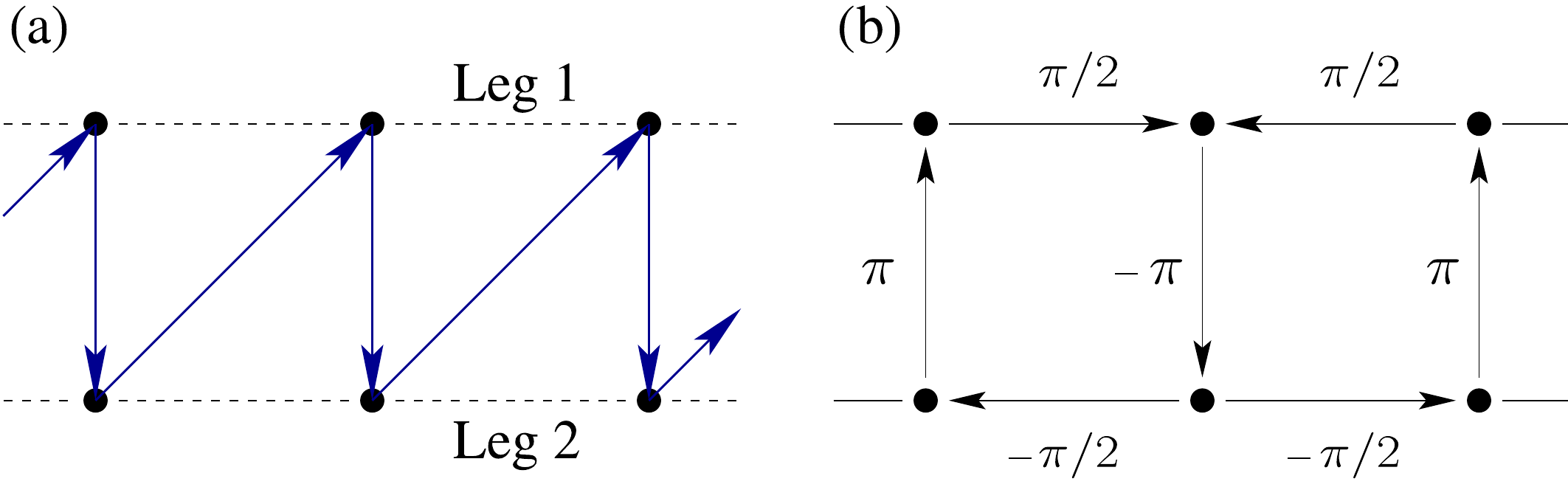}
\caption{ (Color online)
          (a) Zigzag path  used for the Jordan-Wigner transformation, Eq.\  (\protect\ref{eq:jordanwigner}), 
          (b) gauge transformation, leading to Eq.\ (\protect\ref{Ham-pi}).}
\label{fig:lattice-meanfield}
\end{figure}

The definition of the Jordan-Wigner transformation for the ladder topology
relies on the choice of a path labeling different sites. With the choice
shown in Fig.\ \ref{fig:lattice-meanfield}a it reads:

\begin{eqnarray}
 {S}^{z}_{1,i}&=&{n}_{1,i}-\tfrac{1}{2}
\quad
 {S}^{z}_{2,i}={n}_{2,i}-\tfrac{1}{2}
\nonumber\\
 {S}^{+}_{1,i}&=& {c}^{\dagger}_{1,i}
\exp\big{(}-\ii\pi\sum^{i-1}_{l=1}( {n}_{1,l}+ {n}_{2,l})\big{)}
\nonumber\\
 {S}^{+}_{2,i}&=& {c}^{\dagger}_{2,i}
\exp\big{(}-\ii\pi(\sum^{i}_{l=1} {n}_{1,l}+\sum^{i-1}_{l=1} {n}_{2,l}
\big{)})
\label{eq:jordanwigner}
\end{eqnarray} 
where $ {n}_{\alpha,i}= {c}^{\dagger}_{\alpha,i} {c}_{\alpha,i}$
is the density operator at site $i$ with leg index $\alpha=1,2$.
$ {c}^{\dagger}_{\alpha,i}$ and $ {c}_{\alpha,i}$ are spinless fermionic
creation and annihilation operators which  satisfy the  anticommutation rules
$ \left\{ {c}_{\alpha,i}, {c}^{\dagger}_{\beta,j}\right\}
=\delta_{ij}\delta_{\alpha\beta}.
$  
%
%
Before proceeding to the details of the mean-field calculations below, 
let us summarize the results obtained within this approach. The 
ground state of the asymmetric ladder is characterized by spinless fermions 
circulating around a plaquette thereby 
allowing for a $\pi$-flux phase as solution of the mean-field
equations. This leads to a spin gap $ \Delta \propto  |J_{\perp}| \cos (\theta/2)$ 
for $  |J_{\perp}| \alt   J_{\|}\cos (\theta/2)$. Such regime is unavailable for 
the single-pole system at $\theta =\pi$  in which case we obtain an (indirect) spin gap
$\Delta \propto J_{\bot}^2 /J_{\|}$. The mean-field also predicts a smooth crossover between these two regimes.

Our analysis is rather standard and similar to Ref.\ \onlinecite{Azzouz94}. After application
of the Jordan-Wigner transformation, the Heisenberg Hamiltonian of
Eq.\ \eqref{eq:SSHL-hamilton1} 
can be written as:

\begin{eqnarray}
 {\mathcal{H}}
&=& 
 -\frac{J_{\|}}{2}\sum^{L}_{i=1}
    \left(  (\hat A^{(1)}_{i} )^2  - \hat A^{(1)}_{i} e^{\ii\pi\hat{n}_{i,2}} \right )  
    \nonumber\\
& & - \frac{J_{\|}}{2} \cos^{2}\left(\tfrac{\theta}{2}\right)
 \sum^{L}_{i=1}
   \left(   (\hat A^{(2)}_{i} )^2  - \hat A^{(2)}_{i} e^{\ii\pi\hat{n}_{i+1,1}} \right) ,
  \nonumber\\
 & &+ \frac{J_{\bot}}{2}\sum^{L}_{i=1} (\hat{B}_i^2 - \hat{B}_i)    \nonumber
\end{eqnarray}  
Here, we have defined
\begin{equation}
\hat  A^{(\alpha)}_{i}= {c}^{\dagger}_{\alpha,i} {c}_{\alpha,i+1} +h.c.
\,\, ,\quad
\hat{B}_{i}={c}^{\dagger}_{1,i}{c}_{2,i} + h.c. 
\end{equation} 
We restrict ourselves below to  a phase with zero magnetization,
$\langle{S}^{z}_{\alpha,i}\rangle=0 $, which corresponds to 
$ \langle  {n}_{\alpha,i}\rangle=\frac{1}{2}$.
To proceed further we will replace $ {n}_{\alpha,i} $ by its mean value.
Although this simplification cannot be rigorously justified \cite{Aristov1998}
and more elaborate treatments are possible in a Jordan-Wigner
approach \cite{Nunner04}, we show below that it qualitatively reproduces the
results available by more sophisticated methods. 

The replacement $e^{\ii\pi\hat{n}_{\alpha,i+1}} \to i $ defines a spiral $U(1)$ phase shift for fermions 
on each leg. The relative phase between these spirals demarcate two qualitatively different situations. 
In one case the phase flux $\Phi$ through each plaquette is zero and in another case it is equal to  $\pi$. 
Using the gauge transformation, we can reduce the  $\Phi$-flux phase effective Hamiltonian to the form

\begin{equation}
\begin{aligned}
 \mathcal{H}_\Phi
=& 
 -\frac{J_{\|}}{2}\sum_{l}
    \left(  (\hat A^{(1)}_{l} )^2  + \hat A^{(1)}_{l} \right )  
   \\
 & - \frac{J_{\|}}{2} \cos^{2}\left(\tfrac{\theta}{2}\right)
 \sum_{l}
   \left(   (\hat A^{(2)}_{l} )^2  + \hat A^{(2)}_{l}   \right) ,
 \\
 & + \frac{J_{\bot}}{2}\sum_{l} (\hat{B}_l^2 +  e^{i \Phi l} \hat{B}_l)    
\end{aligned}
 \label{Ham-pi}
\end{equation}  
with a checkerboard character of coupling in the $J_{\bot}$ channel happening in the $\pi$-flux case, $\Phi = \pi$.  
On the other hand, the $0$-flux case is characterized by a uniform $J_{\bot}$ coupling, $e^{i \Phi l} \to 1$ in Eq.\ (\ref{Ham-pi}) . 

We then use the mean-field decoupling, 
\[ 
\langle \hat A^{(j)}_{l} \rangle =   A^{(j)} , \quad
\langle \hat B_{l} \rangle = e^{i \Phi l} B 
\]
for the quadratic part of the Hamiltonian (\ref{Ham-pi}), where the special property 
$\hat B_l^3 = \hat B_l$, $(\hat A^{(j)}_{l})^3 = \hat A^{(j)}_{l}$ should be taken into account. 
The remaining Hamiltonian is quadratic in fermions and its 
spectrum is easily found. 
It can be shown that the $\pi$-flux phase provides a lower ground state energy, and hence we focus on 
this phase from now on. 
From the consistency equations below it can be shown that $A^{(1)} =  A^{(2)} = A$, and we can use 
this observation to simplify our subsequent formulas. 

The spectrum has two bands, 
\begin{eqnarray}
\varepsilon^{(\pm)}_{q} &=& \tfrac 12 
J_{\|} (1+A) \sin^2 (\tfrac\theta2 )  \cos q \pm   \tfrac 12  E_q ,
\label{eq:MFspectrum}
\\
E_q & = &
\sqrt{
(J_{\|} (1+A)(1+\cos^2 (\tfrac\theta2 ) ) \cos q )^2+ 
J^{2}_{\perp} (1+B) ^2
}, \nonumber
\end{eqnarray}
and this dispersion should be combined with the consistency conditions
 
\begin{eqnarray}
A &=&
     \frac{1}{L}  \sum_{{ q}}  \frac{J_{\|}(1+A)(1+\cos^2 (\tfrac\theta2 ))\cos^{2} q}{E_q },
      \\
B &=& 
\frac{1}{L}
      \sum_{{ q}}\frac{J_{\bot}(1+B)}
      {2 E_q},
\end{eqnarray}
where the thermodynamic limit $\frac{1}{L}  \sum_{{q}}  \to \int dq/2\pi  $ is assumed.  

In the limit $J_\perp \to 0 $ we have 
$A \simeq 2/\pi$ and 
\[
B \simeq  \frac {(J_\perp/J_\| ) \ln (J_\|/J_\perp c_{1})}{(1+\pi/2)(1+\cos^2 (\tfrac\theta2 ) )}.
\]
with $c_{1} \sim 1$.  
   
At half-filling, corresponding to a vanishing total magnetization in the spin 
language,  a \emph{direct} gap  is given by $\varepsilon^{(+)}_{q} - \varepsilon^{(-)}_{q} = E_{q}$ at $q=\pi/2$ or 
\begin{equation}
\begin{aligned}
\Delta_0 = &\left|J_{\bot}\left(1+B\right) \right| , \\
\simeq & \left| J_{\bot} \right| \left(1+ O( (J_\perp/J_\| ) \ln (J_\|/J_\perp))\right) 
\end{aligned}
\label{eq:MFgap1}
\end{equation}
Hence, this simple mean-field approach is consistent (including logarithmic corrections) 
with bosonization~\cite{Shelton1996} and quantum
Monte Carlo simulations~\cite{Larochelle04} for which a spin gap of the form Eq.\ (\ref{eq:MFgap1}) is found in 
the weak coupling limit.

The \emph{indirect} gap, which is the minimum excitation energy in 
the spin system at zero temperature, is defined as 
$\Delta = \min_q \varepsilon^{(+)}_{q} - \max_k \varepsilon^{(-)}_{k} = 2 \min_q \varepsilon^{(+)}_{q}$. 
The qualitative form of the spectrum (\ref{eq:MFspectrum}) at $\theta \simeq \pi$
is depicted  in Fig.\ 2 of  Ref.\  \onlinecite{Kiselev05a}. 
For small $J_{\bot}$, a flat band is apparent reflecting the macroscopic
degeneracy of the model at $\theta = \pi$ and $J_{\bot} = 0$. This leads to a
dense spectrum of particle-hole excitations at low energies.
A straightforward calculation shows that 
\begin{equation}
\Delta= \Delta _0 \frac{2 \cos\tfrac\theta2}{1+ \cos^2\tfrac\theta2} 
\end{equation} 
at $\cos\tfrac\theta2 \sim 1$ and small enough $J_\perp$. When $\cos \tfrac\theta2 \to 0$, the domain of linear 
dependence of $\Delta$ on $J_\perp$ disappears 
and we have the quadratic law. Expressing energies in units of $\epsilon_0 = J_\|(1+A)(1+ \cos^2\tfrac\theta2 )$,  
we obtain for small $J_\perp/\epsilon_0 , \cos\tfrac\theta2 \ll 1$ : 
\begin{eqnarray}
\Delta & \simeq&  2 J_\perp   \cos\tfrac\theta2 , \quad J_\perp < 2 \epsilon_0 \cos\tfrac\theta2
\nonumber \\
&\simeq& J_\perp ^2 /(2\epsilon_0 ) + 2 \epsilon_0    \cos^2 \tfrac\theta2, 
\quad J_\perp  > 2 \epsilon_0 \cos\tfrac\theta2  .
\label{gapMFT} 
\end{eqnarray}
A linear regime for the indirect gap occurs for incommensurate $q$ in the above 
expression $\min_q \varepsilon^{(+)}_{q}$, whereas 
the quadratic regime in (\ref{gapMFT}) corresponds to the difference 
$\varepsilon^{(+)}_{q=\pi } - \max_k \varepsilon^{(-)}_{k=0}$, i.e.\ commensurate wave-vector $\pi$ of the 
particle-hole excitation.

\section{magnons for long-range interaction}
\label{sec:app1}

We use the Dyson-Maleyev representation of spin operators. In a predominantly AF 
situation, we write   
\begin{equation}
\begin{aligned} 
S^z_l =&  (s - a_l^\dagger a_l)(-1)^l \\ 
S^x_l =& \sqrt{s/2}(a_l^\dagger + a_l - a^\dagger_l a_l^2/(2s) ) \\
S^y_l =& i\sqrt{s/2}(a_l^\dagger - a_l +a^\dagger_l a_l^2/(2s))(-1)^l
\end{aligned}
\end{equation}
For spins in different sublattices we have 
\begin{eqnarray*}
{\bf S}_1 {\bf S}_2 &=&
 -s^2-s + s (a^\dagger_1 + a_2)(a^\dagger_2 + a_1) + \ldots
\end{eqnarray*}
and for spins in the same sublattice :
\begin{eqnarray*}
{\bf S}_1 {\bf S}_3 &=&
 s^2 - s (a^\dagger_1 - a^\dagger_3)(a_1 - a_3) + \ldots
\end{eqnarray*}
Adopting the FM interaction $\tilde V(r)$ within one sublattice and
AFM interaction $V(r)$ between sublattices, we
come to the linearized Hamiltonian :
\begin{equation}
H = s \sum_k (2 a^\dagger_ka_k [V(0) + \tilde V(0)-\tilde V(k)]
+V(k)[ a^\dagger_k a^\dagger_{-k}+ h.c. ])
\end{equation}
Notice that we have $V(k+\pi) = -V(k)$ and  $\tilde V(k+\pi) = \tilde V(k)$.
 To stress the similarity to acoustic phonons, the last equation can also be rewritten as
\begin{equation}
H = s \sum_k (g_{k+\pi} P_k P_{-k}
+ g_{k} Q_k Q_{-k}  )
\end{equation}
where
\begin{equation}
\begin{aligned} 
P_k=& (a^\dagger_k + a_{-k})/\sqrt{2} \\
Q_k=& i (a^\dagger_k - a_{-k})/\sqrt{2} \\
g_k =& [V(0)-V(k) + \tilde V(0)-\tilde V(k)]
\end{aligned}
\end{equation}
so that canonical commutation relations hold, $[P_k, Q_q] = i \delta(k+q)$.
For $k\simeq 0 $ we have $g_k \propto k^2$, $g_{k+\pi} \simeq 2 V(0)$
 the spectrum $\omega(k) = 2s \sqrt{g_kg_{k+\pi}}
\sim k$ is linear for small $k$. The analogy with acoustic
phonons is incomplete, because in the second magnetic Brillouin zone (close to the point $k\simeq \pi$)
the role of $P_k$ and $Q_k$ is reversed with respect to the form of their correlation functions.

For the dynamical susceptibility, $\chi^{xx}$,
the representation in terms of $a$ is the same for both sublattices and
we have $S^x_k = \sqrt{s}P_k$.
The equations of motion read 
\begin{equation}
\begin{aligned} 
\partial_t P_k =& 2 s  g_{k} Q_k \\
\partial_t Q_k =& - 2 s  g_{k+\pi} P_k
\end{aligned}
\end{equation}
and hence
\begin{equation}
\chi^{xx}(k,\omega) = s \frac{2 s  g_{k}}{\omega(k)^2 - \omega^2}
\end{equation}
 
For the nearest neighbor interaction $J$ this formula simplifies to:
\begin{equation}
S\frac{2SJ(1-\cos q)}{\Omega_m^2+ (2SJ\sin q)^2}.
\end{equation}



\end{document}